\documentclass[11pt,a4paper,bookmarks=false]{article}
 \pdfoutput=1
\usepackage{jheppub}
\usepackage{bm,longtable,enumerate}
\usepackage{oldgerm}
\usepackage{indentfirst} 
\usepackage{setspace}
\usepackage{multicol}
\usepackage{multirow}
\usepackage{verbatim}
\usepackage[utf8]{inputenc}
\usepackage{mathrsfs}
\usepackage{amsmath}

\newcommand{\expva}[1]{\langle #1 \rangle}

\title{
Third-order relativistic hydrodynamics: dispersion relations and
transport coefficients of a dual plasma}
\author[\,a]{Saulo M. Diles,} 
\author[\,b,\,c]{Luis A. H. Mamani,}
\author[\,b]{Alex S. Miranda,}
\author[\,c]{and Vilson T. Zanchin}
\affiliation[a]{
Faculdade de F\'\i sica, Universidade Federal do Par\'a, Campus Salin\'opolis,
Rua Raimundo Santana Cruz S/N,
68721-000 Salin\'opolis, Par\'a, Brazil}
\affiliation[b]{ 
Departamento de Ci\^encias Exatas e Tecnol\'ogicas, 
Universidade Estadual de Santa Cruz, Rodovia Jorge Amado, km 16, 45650-000 Ilh\'eus, Bahia, Brazil}
\affiliation[c]{
Centro de Ci\^encias Naturais e Humanas, 
Universidade Federal do ABC, Avenida dos Estados 5001, 
09210-580 Santo Andr\'e, S\~ao Paulo, Brazil}
\emailAdd{smdiles@gmail.com}
\emailAdd{luis.mamani@ufabc.edu.br}
\emailAdd{asmiranda@uesc.br}
\emailAdd{zanchin@ufabc.edu.br}

\abstract{Hydrodynamics is nowadays understood as an effective field theory that describes the dynamics of the
long-wavelength and slow-time fluctuations of an underlying microscopic theory. In this work we extend the
relativistic hydrodynamics to third order in the gradient expansion for neutral fluids in a general curved
spacetime of $d$ dimensions. 
We find 58 new transport coefficients, 19 due to third-order scalar corrections and 39 due to tensorial corrections.
In the particular case of a conformal fluid, the number of new transport coefficients is reduced to 19,
all of them due to third-order tensorial corrections.
The dispersion relations of linear fluctuations in the third-order
relativistic hydrodynamics is obtained, both in the rest frame of the fluid and in a general moving frame.
As an application we obtain some of the transport coefficients of a relativistic conformal fluid in three dimensions
by using the AdS/CFT correspondence. These transport coefficients are extracted from the dispersion relations
of the linear fluctuations. The gravity dual of the fluctuations in this conformal fluid
is described by the gravitational perturbations of four-dimensional anti-de Sitter black branes,
which are solutions of the Einstein equations with a negative cosmological constant. 
To find the hydrodynamic quasinormal modes (QNMs) of the scalar sector
we use the SUSY quantum mechanics of the gravitational perturbations
of four-dimensional black branes.  
Such a symmetry allows us to find the wavefunction of the scalar (or sound) sector in the hydrodynamic limit
directly from the wavefunction of the vector (or shear) sector, which is usually easier to be found
because the perturbation wave equations for the vector sector are much simpler than the ones for the scalar sector.
}

\keywords{ Relativistic Hydrodynamics, AdS/CFT, Black holes, Classical Theories of Gravity, Quasinormal Modes}
\arxivnumber{}

\begin{document}
\maketitle

\section{Introduction}
\label{introd}

The very unique properties of the quark-gluon plasma claim for the development of 
theoretical physics. In the last century it was speculated that, for very high 
temperatures, quarks and gluons would behave like free particles as a consequence of the asymptotic freedom
\cite{Cabibbo:1975ig}. In this century  
it has become possible to reach such very high temperatures in particle accelerators like the RHIC and LHC, where  
the quark-gluon plasma (QGP) has been produced in heavy ion collisions since 2005. Despite
those initial  expectations, quarks and gluons are far from being free in the QGP; rather, they are  strongly coupled and the pertubative QCD is unable to give reliable predictions for this system \cite{Busza:2018rrf}. Many challenges need to be overcome to find a complete theoretical description of the QGP since, besides  
the very high temperatures and strongly coupled interactions involved in the process, the quarks and gluons are confined so that direct measurements in QGP are forbidden and only the final hadronic states are observable \cite{Gyulassy:2004vg}. 
Such a difficulty may be circumvent by following the 1950's Landau proposal to make use of hydrodynamics to describe the collective behaviour of fundamental matter \cite{Belenkij:1956cd}. In fact, such a proposal has been successfully applied in the description of heavy ion collisions \cite{Luzum:2008cw,Gyulassy:2004vg,Kapusta:2011gt,deSouza:2015ena,Busza:2018rrf}. Therefore, the direct application of fluid dynamics
to describe the evolution of QGP is a strong motivation to look for further developments on the theory of relativistic hydrodynamics.

The relativistic version of the Navier-Stokes equation has a well-known causality problem
\cite{Hiscock:1985zz}. It has a mode that satisfies a partial differential equation of the heat kind whose Green function allows for instantaneous propagation. In this sense the relativistic Navier-Stokes equations are inconsistent with special relativity itself and a more refined treatment is required.
The first refinement comes by considering two additional terms, each containing one extra derivative, first implemented for non-relativistic fluids \cite{Muller:1967zza} and then extended to the relativistic ones
\cite{Israel:1976tn, Israel:1979wp}. The additional terms come with new transport coefficients, the relaxation times, and solve the causality problem, providing a consistent theory of viscous relativistic fluids
known as the Muller-Israel-Stuart (MIS) theory.  

It turns out that the additional terms included in the MIS theory to solve the causality problem do not represent the most general setup that should include first and second derivatives of the fundamental degrees of freedom of the fluid. In fact, the relativistic version of the Navier-Stokes equation, as well as the ideal fluid equations, can be expressed by the divergenceless of an energy-momentum tensor, suggesting that hydrodynamics can be formulated by fixing the general form of this energy-momentum tensor. This strategy has been used in \cite{Baier:2007ix} where the complete theory of conformal fluid dynamics involving terms up to second order in derivatives of the energy-momentum tensor is obtained (see also Ref.~\cite{DiNunno:2017obv}, where the Bjorken flow was investigated). For non-conformal fluids the same procedure of fixing the most general energy-momentum tensor involving up to second-order derivatives was used in \cite{Romatschke:2009kr}, thus generalizing the MIS theory. This way of constructing the theory of fluid dynamics by fixing the most general effective energy-momentum tensor up to some order in derivatives is known as the gradient expansion.  

Now, by means of the AdS/CFT correspondence \cite{Maldacena:1997re}, the gradient expansion
may be connected to gravitational perturbations on the anti-de Sitter (AdS) side of the duality. 
In fact, we learned from the AdS/CFT that the energy-momentum tensor in a conformal quantum field theory 
is dual to the background metric of an asymptotically AdS spacetime.
As a consequence, linear fluctuations around a thermal-equilibrium state in the field theory 
correspond to linear gravitational perturbations on an AdS black hole.  
Furthermore, the spectra of black-hole quasinormal modes (QNMs) \cite{Berti:2009kk} are
characterized as linear perturbations with infalling boundary condition at the horizon, 
which is equivalent to consider retarded Green's functions in the dual field 
theory. The relation between the poles of retarded Green's functions and the QNMs requires that the condition imposed 
on the metric fluctuations  at the AdS boundary must be of the Dirichlet type. With 
these boundary conditions the frequencies of the QNMs are complex numbers, whose imaginary parts 
may be interpreted as thermalization times in the dual field theory.

Still regarding the linear fluctuations of a black-hole metric, an underlying symmetry arises
in four-dimensional spacetimes, which is similar to the supersymmetric (SUSY) quantum mechanics
\cite{Witten:1981nf,Witten:1982df,Cooper:1994eh} and relates the axial (odd) and polar (even)
sectors of the gravitational perturbations. Such a symmetry was first noticed
in the context of the transformation theory developed by Chandrasekhar \cite{Chandrasekhar:1983}
and applied to the study of the metric perturbations of Schwarzschild black holes in asymptotically flat spacetimes 
(see for instance Refs. \cite{Burgess:1984ti,Mellor:1989ac,Leung:1999}). 
In the linearized gravity, the SUSY relation represents the gravitational counterpart of
the electric/magnetic duality of the vacuum Maxwell equations \cite{Bakas:2008gz, Henneaux:2004jw}. 
Additionally, the SUSY quantum mechanics of perturbations also occurs in the presence of a
(positive or negative) cosmological constant. In particular, it has been shown to hold in four-dimensional
asymptotically AdS spacetimes  \cite{Bakas:2008zg, Miranda:2014vaa}.

Asymptotically anti-de Sitter black-hole like objects play a very important role in
various applications of the gauge/gravity duality.
A particularly interesting case within this family is the four-dimensional black brane 
\cite{Lemos:1994fn,Lemos:1994xp,Huang:1995zb,Cai:1996eg,Lemos:1995cm}. The fact that makes this solution
important for applications in AdS/CFT is the non-spherical topology of the horizon, which may be planar
($ \mathbb{R} \times \mathbb{R}$), cylindrical ($ \mathbb{ S}^1 \times \mathbb{R}$) or toroidal
($\mathbb{S}^1\times \mathbb{S}^1$). The gravitational and electromagnetic perturbations 
of these objects have been extensively studied in the last two decades \cite{Cardoso:2001vs,Miranda:2005qx,Miranda:2007bv,Miranda:2008vb,Morgan:2009pn,Morgan:2013dv,deOliveira:2018jhc}.
In Ref.~\cite{Miranda:2014vaa} the authors studied the gravitoelectromagnetic 
perturbations of charged black branes and extended the Chandrasekhar's
transformation theory to the presence of sources. They found a SUSY
quantum-mechanical relation between the vector (or shear) and the scalar (or sound)
sectors of these perturbations. It is worth mentioning that one of the main consequences
of this symmetry  is that the problem of finding the QNMs can be greatly reduced. For example, 
we might focus on the vector (or shear) sector to find the QNM wavefunctions, and
then use the SUSY relation to find the solutions of the scalar (or sound) 
sector, whose perturbation equations are in general more complicated to be solved.
Such a fact is used along this work.

This work is organized as follows.
Section.~\ref{Sec:NonConformalFluids} is dedicated to the study
of the general case of non-conformal fluids in $d$-dimensional spacetimes. We use a systematic 
procedure to expand the energy-momentum tensor up to third order 
in the gradient expansion. In Section~\ref{Sec:ConformalFluids} we 
deal with conformal fluids. The conformal symmetry is implemented via minimal coupling
what allows us to obtain third order conformal corrections in a direct way.
Section~\ref{Sec:DispersionRelations} is devoted to study linearized fluctuations of the fluid and to find the 
dispersion relations as a functions of a reduced set of transport coefficients
for either non-conformal and conformal fluids. In 
Section~\ref{Sec:Application} we analyze an application of the general results 
obtained in previous sections, focusing on the gravitational perturbations 
of plane-symmetric static black holes, where we find the dispersion relations 
in the hydrodynamic limit, i.e., in the low energy and long wavelength limit,
of the vector (transverse or shear) sector of the perturbations. Then, we use the SUSY relation to find the dispersion
relations of the scalar (longitudinal or sound) sector
in the hydrodynamic limit. We also include a discussion on different 
kind of variables that might be used to write the gravitational 
perturbations. At the end, we compute the energy transport 
coefficients arising up to third order in the gradient expansion 
of the conformal fluid, which is dual to the plane-symmetric static black hole.
We present the conclusion and final remarks in Section~\ref{Sec:Conclusion}.
Additional material are left in Appendices \ref{Sec:AppendixA} and \ref{Sec:AppendixB}.
The explicit verification of some equivalences among third-order structures containing
transverse derivatives of second order is presented in Appendix \ref{Sec:AppendixC}.

\section{Relativistic hydrodynamics}

\label{Sec:NonConformalFluids}
\subsection{The basic formulation}

The purpose here it to construct order by order the effective theory that describes the dynamics of a relativistic one-component and uncharged fluid.
This is done by looking for corrections to the ideal fluid dynamics
in the form of gradients of the fundamental degrees of freedom.
The local fundamental variables of a fluid are the energy density $\varepsilon$, pressure $p$, entropy density $s$, temperature $T$,
and the vector field $u^\mu$ corresponding to the flow velocity of a fluid element. In order to account for gravitational
effects we assume the metric field to be a local degree of freedom,
with the restriction in this aspect to torsion-free geometries, so that $g_{\mu\nu}=g_{\nu\mu}.$
The relativistic equations for an ideal fluid are obtained
from the divergenceless condition $\nabla_\mu T^{\mu\nu}_{\mbox{\scriptsize{ideal}}}=0$ on the energy-momentum tensor
\begin{equation}
T^{\mu\nu}_{\mbox{\scriptsize{ideal}}}= \varepsilon u^\mu u^\nu + p \Delta^{\mu\nu},
\label{energia_momento_ideal}
\end{equation}
where $\Delta^{\mu\nu}=g^{\mu\nu}+u^\mu u^\nu$ is the projector onto the
hypersurface orthogonal to the velocity $u^\mu$, and we adopt the metric signature $(-,+,+,...,+).$
The energy-momentum tensor of an ideal fluid does not contain derivatives of the local degrees of freedom,
and its divergence-free equation $\nabla_\mu T^{\mu\nu}_{\mbox{\scriptsize{ideal}}}=0$ 
leads to 
\begin{equation}\begin{aligned}
& u^\mu\nabla_\mu \varepsilon + (\varepsilon+p)\nabla_\mu u^\mu = 0,\\ & \left(\varepsilon+p\right) u^\mu\nabla_\mu u^\nu 
+\Delta^{\mu\nu}\nabla_\mu p =0.
\end{aligned} \label{idealeq}
\end{equation}
Alternatively, the last equations for the perfect fluid can be derived from a Lagrangian formulation,
as it has been done in Ref. \cite{Taub:1954zz}, 
where the Einstein equations sourced by an ideal fluid were also obtained.  
However, for viscous relativistic fluids, a variational formulation to derive the equations governing the dynamics of the system is still under development (see, for instance, Refs.
\cite{Kovtun:2014hpa,Grozdanov:2013dba,Haehl:2014zda,Haehl:2015pja,Haehl:2015foa,Haehl:2015uoc,Crossley:2015evo,Glorioso:2017fpd,Haehl:2018uqv,Haehl:2018lcu,Jensen:2017kzi,Jensen:2018hse}). 
This is not a problem here, since such a formulation is not necessary, the construction of the hydrodynamics is done at the level of the energy-momentum tensor conservation, assuming that $\nabla_\mu T^{\mu\nu}=0$.

\subsection{Gradient expansion}

The starting point to construct the gradient expansion is the ideal fluid, which has the energy-momentum tensor $T^{\mu\nu}_{\mbox{\scriptsize{ideal}}}$ given by \eqref{energia_momento_ideal}.
In the context of a perturbative expansion, the ideal fluid corresponds to the zero-order (unperturbed) system, and the perturbations
involving derivatives are responsible by the introduction of dissipative effects. 
The energy-momentum tensor has its own symmetry properties: it is a divergenceless second-rank symmetric tensor, so that all corrections to
\eqref{energia_momento_ideal} should preserve these properties. 

The gradient expansion corrects $T^{\mu\nu}_{\mbox{\scriptsize{ideal}}}$ by terms involving some number of derivatives of the degrees of freedom:
first-order corrections involve terms with only one derivative (first-order derivative terms), second-order corrections involve
either terms with two derivatives (second-order derivative terms) and/or the product of two terms, each of them
containing one derivative (quadratic first-order derivative terms), and so on.
Schematically the energy-momentum tensor of the fluid at order $n$ is given by
\begin{equation}
T_{\scriptscriptstyle{(n)}}^{\mu\nu} =  T^{\mu\nu}_{\mbox{\scriptsize{ideal}}}+
\sum_{i=1}^n\Pi_{\scriptscriptstyle{(i)}}^{\mu\nu},
\end{equation}
where each $\Pi_{\scriptscriptstyle{(i)}}^{\mu\nu}$ represents all possible corrections at order $i$ in derivatives of the fundamental variables.
The most general form for $\Pi_{\scriptscriptstyle{(i)}}^{\mu\nu}$ can be decomposed
into scalar, vector and tensor components as
\begin{equation} 
\Pi^{\mu\nu}_{\scriptscriptstyle{(i)}} =\mathcal{E}_{\scriptscriptstyle{(i)}} u^\mu u^\nu +
\mathcal{P}_{\scriptscriptstyle{(i)}}\Delta^{\mu\nu} +\left(q_{\scriptscriptstyle{(i)}}^\mu u^\nu +
q_{\scriptscriptstyle{(i)}}^\nu u^\mu\right) + t_{\scriptscriptstyle{(i)}}^{\mu\nu}, 
\label{corrections}
\end{equation}
where $q_{\scriptscriptstyle{(i)}}^\mu$ are transverse vectors $\left(u_\mu q_{\scriptscriptstyle{(i)}}^\mu=0\right)$
and $t_{\scriptscriptstyle{(i)}}^{\mu\nu}$ are transverse, symmetric and traceless (TST)
tensors\footnote{The TST part of a general second-rank tensor $\mathcal{T}^{\mu\nu}$ is given by 
$ 
 \mathcal{T}^{\expva{\mu\nu}}= \dfrac{1}{2}\Delta^{\mu\alpha}\Delta^{\nu\beta}(\mathcal{T}_{\alpha\beta}+\mathcal{T}_{\beta\alpha}) - \dfrac{1}{d-1} \Delta^{\mu\nu}\Delta^{\alpha\beta}\mathcal{T}_{\alpha\beta}.
$ }.

In the Landau frame one identifies the longitudinal projection of the energy-momentum tensor with
the energy density, $u_\mu T_{\scriptscriptstyle{(n)}}^{\mu\nu}=-\varepsilon u^\nu$, and this requires
the corrections to be transverse: $u_\mu \Pi_{\scriptscriptstyle{(i)}}^{\mu\nu}=0$. 
Such a choice of frame eliminates the corrections of the energy density
$\mathcal{E}_{\scriptscriptstyle{(i)}}=0$, so that the scalar gradients will change only the pressure,
and eliminates all vector corrections $q_{\scriptscriptstyle{(i)}}^\mu=0.$
The freedom of frame choice effectively reduces the number of transport coefficients, since one can always
choose the Landau frame, which forces 
half of the scalar and all vector corrections to $T^{\mu\nu}_{\mbox{\scriptsize{ideal}}}$ to vanish.  
In this section we focus on finding the set of transport coefficients and we will discuss only scalar and tensor structures. 
Our procedure can also be applied to non-conformal vector structures 
and we include the list of 
independent correction terms of vector type in Appendix \ref{Sec:AppendixB}. 

In order to implement  corrections of the ideal fluid dynamics we look for the complete set of admissible terms, 
which should respect the underlying symmetries of the system.  
We are interested here in the description of a very general system; in particular, we consider the
presence of the gravity field so that fluid dynamics is
formulated in a curved background spacetime. The presence of the gravity is introduced in the gradient expansion by using the covariant derivative, e.g., $\nabla_\mu u^\nu = \partial_\mu u^\nu + \Gamma^{\nu\,}_{\;\mu\sigma}u^\sigma$, where $ \Gamma^{\nu\,}_{\;\mu\sigma}$ stands for the Christoffel coefficients. Here we assume that the background geometry is torsion-free. 

The fluid dynamics is characterized by its constitution, thermodynamics, background geometry and flow. We restrict the analysis to uncharged fluids in local thermodynamic equilibrium, which reduces the independent degrees of freedom  
to be the velocity field $u^\mu$, the logarithm of the entropy density\footnote{Notice that the energy density $\varepsilon$ and the pressure $p$ that appear explicitly in the ideal-fluid energy-momentum tensor, and the local temperature $T$, are related to $\ln s$ by thermodynamic equilibrium equations.} $\ln s$,
and metric $g_{\mu\nu}$. Corrections to the ideal fluid are build with gradients of these degrees of freedom. Moreover, zeroth-order equations of motion impose additional relations\footnote{These relations are equalities in the sense that their corrections require higher-order gradients.}
\begin{equation}
 D \ln s = -\nabla\cdot u,\qquad Du^\mu = -c_s^2\nabla_{\perp}^\mu \ln s, 
\end{equation}
where $c_s$ is the speed of sound in the fluid, and we have defined the transverse derivative
by $\nabla_{\perp}^\mu = \Delta^{\mu\nu}\nabla_\nu$ and the longitudinal (or time)
derivative as $D = u^\mu \nabla_\mu$.
The longitudinal derivative of the flow velocity is equivalent to the entropy gradient, and here we choose to use only the transverse gradient $\nabla_{\perp}^\mu u^\nu$ that is a second-rank tensor and can be decomposed as
\begin{equation}
 \nabla_{\perp}^\mu u^\nu =  \sigma^{\mu\nu} + \Omega^{\mu\nu} - \frac{\Theta}{d-1} \Delta^{\mu\nu}, \label{eqideal}
\end{equation}
where $\sigma^{\mu\nu} = \nabla_{\perp}^{\langle \mu} u^{\nu\rangle}$ is the symmetric and traceless shear tensor, 
$\Omega^{\mu\nu} = \nabla_{\perp}^\mu u^\nu-\nabla_{\perp}^\nu u^\mu$  is  the anti-symmetric vorticity tensor, and $\Theta = \nabla_{\perp}\cdot u$ is the scalar expansion. The requirement that velocity field is normalizable, $u^\mu u_\mu=-1$, provides further identities regarding the first-order gradient of the velocity flow, such as $u_\mu\nabla_\nu u^\mu=0$.

The strategy suggested in \cite{Grozdanov:2015kqa} is to build the corrections \eqref{corrections}
by acting with the gradient operator on the fundamental degrees of freedom and contracting
the resulting quantities systematically with the aim to obtain all the independent terms at the corresponding order in the gradient expansion (see Refs.~\cite{Lublinsky:2009kv,Bu:2014sia,Bu:2014ena} for a generalization at linear order).
At each order one takes the terms coming from differentiating $\ln s, u^\mu, g_{\mu\nu}$, contracts in all possible ways, and then uses the equations of motion and other algebraic constraints to eliminate the redundancies. This procedure works fine at fist and second order, but for third-order terms it is not easy to rule out redundancies due to the high number of possibilities, what may lead to mistakes. There is no problem in such a systematic approach, but if one uses it together with the strategy of working with the irreducible gradients as fundamental blocks, instead of the fundamental degrees of freedom, as done, e.g,. in Ref. \cite{Bhattacharyya:2012nq}, one finds out the complete set of possible corrections in a more direct way. The use of irreducible gradients as fundamental structures seems to be a natural way of performing a gradient expansion since one is in fact looking for the irreducible set of higher-order gradients. In the following we work out the proposed strategy and obtain the complete irreducible set of third-order gradients for non-conformal and conformal hydrodynamics.
The obtained results complement those presented in \cite{Grozdanov:2015kqa} by revealing equivalences between the third-order structures for non-conformal fluids and presenting the full set of third-order scalars and vectors for conformal fluids.

\subsection{Revisiting the first- and second-order hydrodynamics}
\label{Sec:FirstSecondHydro}

To find the higher-order corrections of hydrodynamics we take the set of irreducible gradients 
of the fundamental degrees of freedom $\{\nabla_{_\perp}^\mu\ln s,~\Theta,~\Omega^{\mu\nu},~\sigma^{\mu\nu},~R^{~\mu\nu}_{\alpha~~\beta}\}$ to build scalars,
vectors and second rank tensors.
In such a list we find one first-order scalar, one first-order vector,
two first-order tensors of second rank, and one second-order tensor of fourth rank.
The first-order vectors do not need to be considered because we are working in the Landau gauge. Moreover, at first order, there is no contribution from curvature, since it is second order in derivatives. The vorticity tensor is anti-symmetric and hence either its trace and its TST part vanish identically. As the vector $\nabla_{_\perp}^\mu\ln s$ is transverse its projection along $u^\mu$ vanishes and we cannot make a first-order scalar with it. We have one first-order scalar $\Theta$ as well as one TST tensor of second rank $\sigma^{\mu\nu}$ that satisfies all symmetry requirements. Each independent structure we add on the energy-momentum tensor comes with one arbitrary parameter, the transport coefficient, that may depend on the temperature. For the first-order corrections the transport coefficients are the bulk viscosity $\zeta$ and the shear viscosity $\eta$,  and the first-order energy-momentum tensor reads 
\begin{equation} 
\Pi_{\scriptscriptstyle{(1)}}^{\mu\nu}=  - \zeta \Theta\Delta^{\mu\nu} - \eta \sigma^{\mu\nu}.
\end{equation}

First-order hydrodynamics is obtained by the divergenceless of $T_{\scriptscriptstyle{(1)}}^{\mu\nu}
= T_{\mbox{\scriptsize{ideal}}}^{\mu\nu}+\Pi_{\scriptscriptstyle{(1)}}^{\mu\nu}$ and is precisely the
relativistic version of the Navier-Stokes equation. The second law of thermodynamics imposes
additional constraints in the transport coefficients. At first order it requires $\eta>0,~\zeta>0$.

At second order the background curvature, represented by the Riemann tensor, plays an
important role.
The Riemann tensor manifests itself in the second-order derivatives of tensors.
In particular, it is introduced by the
commutation of the covariant derivatives of a vector $\left[\nabla_\mu,\nabla_\nu\right]u^\alpha
= R^\alpha_{~\mu\nu\gamma}u^\gamma$. It establishes that a change in the order of covariant derivatives can be compensated by the addition of the Riemann tensor. Hence, in the task of counting nonequivalent corrections, the order of covariant derivatives is
irrelevant\footnote{The argument presented here for vector fields also holds for any tensor structure; the difference for an $(r,\,s)$ rank tensor is that one needs to add $(r+s)$ terms of the Riemann tensor to compensate each permutation of two covariant derivatives.}
and two structures that differ only by the ordering of the derivatives are in fact redundant.  

Pursuing the task of building second-order corrections, we now turn attention to the second-order scalars. Firstly one notes that the first order scalar $\Theta$ produces two second order scalars, its square $\Theta^2$ and its scalar derivative $D\Theta$. We can also square the first-order vector and the first-order tensors by obtaining
$ (\nabla_{\perp}  \ln s)^2 = \nabla^\mu_{\perp}  \ln s\,\nabla_{\perp \mu}  \ln s$,
$\sigma^2=\sigma^{\mu\nu}\sigma_{\mu\nu}$ and
$\Omega^2 =\Omega^{\mu\nu}\Omega_{\mu\nu}$. Moreover,
one has also to consider the divergence of the first-order vector 
$\nabla^\mu_{\perp }\nabla_{\perp \mu}  \ln s$. Notice, however, that it can be replaced by the remaining structures through the use of Eq.~\eqref{eqideal}. 
Due to the symmetries of the Riemann tensor under index permutations, there are only two nonequivalent scalars
we can build from it, namely, $R = g_{\mu\nu}R^{\mu\nu}$ and $u^\mu u^\nu R_{\mu\nu}$,
where $R_{\mu\nu}$ is the Ricci tensor. The list of independent second-order scalars is given by
\begin{equation}
\begin{aligned}[6]
\mathcal{S}^{^{\mbox{\tiny{2nd}}}}_1 &=\sigma^2,\quad &\mathcal{S}^{^{\mbox{\tiny{2nd}}}}_2 &=
\Theta^2,\quad & \mathcal{S}^{^{\mbox{\tiny{2nd}}}}_3 &
= \Omega^2, &\mathcal{S}^{^{\mbox{\tiny{2nd}}}}_4 &=(\nabla_{\perp}  \ln s)^2, \\
\mathcal{S}^{^{\mbox{\tiny{2nd}}}}_5 &=  D\Theta,\quad &\mathcal{S}^{^{\mbox{\tiny{2nd}}}}_6 &=
R,\quad &\mathcal{S}^{^{\mbox{\tiny{2nd}}}}_7  &=
u^\mu u^\nu R_{\mu\nu}. &
\end{aligned}
\end{equation}

In order to find the second-order tensors, we note that $\sigma^{\mu\nu}$ is a first-order TST tensor and, hence, $\Theta \sigma^{\mu\nu}$ 
and $D\sigma^{\expva{\mu\nu}}$ are second-order TST tensors. Moreover, the symmetric products of two first-order tensors such as $\sigma \sigma,~\sigma\Omega$ and $\Omega\Omega$ are also second-order tensors.
As the Riemann tensor is a second-order correction, its projections and contractions are also second-order corrections and, taking into account its symmetry properties, it results only in two independent structures corresponding to the Ricci and the Riemann tensors projected along the flow. Collecting all the independent second-order tensors we get the following list:
\begin{equation}
\begin{aligned}[6]
\mathcal{T}^{^{\,\mbox{\tiny{2nd}}}}_1 &= \sigma_{\alpha}^{~\langle \mu}\sigma^{\nu\rangle \alpha},
&\mathcal{T}^{^{\,\mbox{\tiny{2nd}}}}_2 & = \sigma_{\alpha}^{~\langle \mu}\Omega^{\nu\rangle \alpha},
& \mathcal{T}^{^{\,\mbox{\tiny{2nd}}}}_3 &= \Omega_{\alpha}^{~\langle \mu}\Omega^{\nu\rangle \alpha},
&\mathcal{T}^{^{\,\mbox{\tiny{2nd}}}}_4 &= \nabla_{_\perp}^{\langle \mu} \ln s\nabla_{_\perp}^{\nu\rangle} \ln s,\\
\mathcal{T}^{^{\,\mbox{\tiny{2nd}}}}_5 &= \Theta\sigma^{\mu\nu},   &\mathcal{T}^{^{\,\mbox{\tiny{2nd}}}}_6 &=
D\sigma^{\expva{\mu\nu}},  &\mathcal{T}^{^{\,\mbox{\tiny{2nd}}}}_7 &= u_\alpha u_\beta R^{\alpha\expva{\mu\nu}\beta},
  &\mathcal{T}^{^{\,\mbox{\tiny{2nd}}}}_8 &=R^{\expva{\mu\nu}},
\end{aligned}
\end{equation} 
where the tensor indexes on left hand side of the equations were omitted.
 
With all scalar and tensor correction terms at hand, we can then write down the most general second-order
correction to the energy momentum-tensor. There are $15$ transport coefficients that we label as $\xi_a$
for the scalar and $\lambda_b$ for tensor corrections, resulting in
\begin{equation}
\Pi_{\scriptscriptstyle{(2)}}^{\mu\nu} =\sum_{a=1}^7 \xi_{a}\,\mathcal{S}^{^{\mbox{\tiny{2nd}}}}_{a}\Delta^{\mu\nu}
+\sum_{b=1}^8 \lambda_b  \left(\mathcal{T}^{^{\,\mbox{\tiny{2nd}}}}_b\right)^{\mu\nu}.   
\end{equation}

The structures just found agree with those 
presented in \cite{Romatschke:2009kr}, the only difference being that we use $\Theta$ instead of $\nabla\cdot u$.
The labeling we adopt here for the transport coefficients is related to that 
employed in \cite{Romatschke:2009kr}
according to the relations $\xi_5=\zeta\tau_\Pi$, $\lambda_6=\eta\tau_\pi$, recovering the relaxation times in the classical literature of relativistic hydrodynamics $\tau_\Pi$ and $\tau_\pi$, and also $\lambda_5=\eta(\tau_\pi+\tau_\pi^*)/(d-1),~\lambda_7=2(\kappa-\kappa^*)$, and
$\lambda_8=\kappa$, while the remaining coefficients perfectly coincide with the notation
employed in~\cite{Romatschke:2009kr}.

\subsection{The third-order hydrodynamics}

Now we use 
the structures defined in the previous section to find the third-order corrections to the energy-momentum tensor of a non-perfect uncharged fluid.
The idea is that, once we have the list of first- and second-order structures, we 
should use it to find the third-order ones. There are two direct ways to build third-order terms:
to take gradients of first- and second-order structures, or to multiply one first-order by one second-order structure.

In order to build scalars by multiplication, we can take the product of 
two scalars, contract two vectors, or contract a tensor with two vectors.
On the other hand, building scalars by differentiating second-order structures
one can take the divergence of second-order vectors or total derivatives of
second-order scalars. These two processes would furnish many equivalent structures,
and so a careful study is demanded. In fact, at third order the geometrical identities
for curvature derivatives plays an important role. For instance, the quantity $u_\mu \nabla_\nu R^{\mu\nu}$ is proportional to $u^\mu\nabla_\mu R$ due to the fact that the Einstein tensor is a divergenceless tensor (i.e., due to the Bianchi identity). Algebraic relations like the anti-symmetry of the vorticity tensor also generates redundancies, eliminating from the list, for example, the scalar and the tensorial structures coming from $\Omega^3$.  After a detailed and careful analysis, we find the irreducible set of third-order scalars to be
\begin{equation}
\begin{aligned}[5]
\mathcal{S}^{^{\mbox{\tiny{3rd}}}}_1 &=\nabla_{_\perp}^2 \Theta,\quad &\mathcal{S}^{^{\mbox{\tiny{3rd}}}}_2 &=
\Theta \nabla_{_\perp}^2 \ln s,\quad &\mathcal{S}^{^{\mbox{\tiny{3rd}}}}_3 &= \sigma^{\mu\nu}\nabla_{_\perp \mu}
\nabla_{\perp \nu}\ln s,\\
\mathcal{S}^{^{\mbox{\tiny{3rd}}}}_4  &= \sigma^{\mu\nu}\nabla_{\perp \mu} \ln s\nabla_{_\perp \nu}\ln s,\quad &
\mathcal{S}^{^{\mbox{\tiny{3rd}}}}_5 &=\nabla_{_\perp \mu} \ln s\nabla_{_\perp \nu}\sigma^{\mu\nu}   ,\quad &\mathcal{S}^{^{3rd}}_6 &=
 \nabla_{_\perp}^\mu\Theta\nabla_{_\perp \mu} \ln s,\\
\mathcal{S}^{^{\mbox{\tiny{3rd}}}}_7  &=\Theta(\nabla_{_\perp} \ln s)^2, \quad &
\mathcal{S}^{^{\mbox{\tiny{3rd}}}}_8  &= \Theta^3,\quad &
\mathcal{S}^{^{\mbox{\tiny{3rd}}}}_9 &= \Theta\Omega^2,\\
\mathcal{S}^{^{\mbox{\tiny{3rd}}}}_{10} &= \Theta\sigma^2\quad &
\mathcal{S}^{^{\mbox{\tiny{3rd}}}}_{11} &=\sigma_{\mu\nu} \sigma^{\mu\eta} \sigma_\eta^{ ~\nu},  ,\quad &\mathcal{S}^{^{\mbox{\tiny{3rd}}}}_{12} &=\sigma_{\mu\nu} \Omega^{\mu\eta} \Omega_\eta^{~\nu},
\\
\mathcal{S}^{^{\mbox{\tiny{3rd}}}}_{13} &= DR,\quad &
\mathcal{S}^{^{\mbox{\tiny{3rd}}}}_{14} &=u^\mu u^\nu D R_{\mu\nu},\quad &\mathcal{S}^{^{\mbox{\tiny{3rd}}}}_{15} &=u^\mu R_{\mu\nu}\nabla_\perp^\mu \ln s,\\
\mathcal{S}^{^{\mbox{\tiny{3rd}}}}_{16} &= \Theta R,\quad & 
\mathcal{S}^{^{\mbox{\tiny{3rd}}}}_{17} &= R_{\mu\nu}\sigma^{\mu\nu}, \quad &\mathcal{S}^{^{\mbox{\tiny{3rd}}}}_{18} &= \Theta u^\mu u^\nu R_{\mu\nu},\\
\mathcal{S}^{^{\mbox{\tiny{3rd}}}}_{19} &=  u^\mu u^\alpha\sigma^{\nu\beta}\mathcal{R}_{\mu\nu\alpha\beta}.
\end{aligned}
\end{equation}

The list obtained here differs from that presented in \cite{Grozdanov:2015kqa} where there
are 23 third-order scalars. In particular the structures $\{\nabla_{\perp \mu}\nabla_{\perp}^\mu
\nabla_{\perp \nu} u^\nu, \nabla_{\perp \mu}\nabla_{\perp \nu}\nabla_{\perp}^\mu u^\nu,\nabla_{\perp \mu}
\nabla_{\perp \nu}\nabla_{\perp}^\mu u^\nu\}$\footnote{Notice that in \cite{Grozdanov:2015kqa}
the authors use lower case Latin indexes to label spacetime components of a tensor.}
are treated as independent structures in \cite{Grozdanov:2015kqa}
but we note that they differ only by the ordering of successive transverse derivatives. The same happens
for the structures $\{\nabla_{\perp}^\mu\ln s \nabla_{\perp \mu}\nabla_{\perp \nu} u^\nu$, $\nabla_{\perp}^\mu\ln s
\nabla_{\perp \nu}\nabla_{\perp \mu} u^\nu\}$. The equivalence of gradients differing only by the order of transverse
derivatives is discussed in Appendix \ref{Sec:AppendixC}.
Moreover, by using the Bianchi identity it may be shown that the scalars
$\{u^\mu \nabla^\nu R_{\mu\nu},~u^\mu\nabla_\mu R \}$ are also equivalent structures.
Then we conclude that there are only three independent scalars from this set of seven,
eliminating 4 transport coefficients and resulting in a reduced list with only 19 scalars.

The tensor structures form a larger list than the scalar structures, since
there are many ways to ``arrange symmetrically"
two-rank tensors involving three derivatives of the fundamental fields.
One can take direct product of one second-order scalar by $\sigma^{\mu\nu}$,
multiply a second-order tensor by $\Theta$, take scalar derivatives of second-order
tensors, or symmetric derivatives of second-order vectors,\footnote{It is important to remark 
that although we work in Landau frame it is useful to have vector structures at hand in order to
find the higher-order scalars and tensors. The vectors are presented in Appendix \ref{Sec:AppendixB}.}
and  also combine in a TST way one first-order vector with one second-order vector. 
After eliminating the equivalent terms, we present below the list of independent third-order structures
by omitting the indexes on the left side:
\begin{equation}
\begin{aligned}[5]
\mathcal{T}^{^{\,\mbox{\tiny{3rd}}}}_1 &=\Omega^2\sigma^{\mu\nu},\quad &\mathcal{T}^{^{\,\mbox{\tiny{3rd}}}}_2
&=\sigma^2\sigma^{\mu\nu},\quad &\mathcal{T}^{^{\,\mbox{\tiny{3rd}}}}_3 &=\Theta^2\sigma^{\mu\nu},\\
\mathcal{T}^{^{\,\mbox{\tiny{3rd}}}}_4 &= \left(D\Theta\right)\sigma^{\mu\nu},\qquad &\mathcal{T}^{^{\,\mbox{\tiny{3rd}}}}_5
&=(\nabla_{_\perp} \ln s)^2\sigma^{\mu\nu}, \quad & \mathcal{T}^{^{\,\mbox{\tiny{3rd}}}}_6
&=\Theta \sigma_\eta^{~\langle \mu}\sigma^{\nu\rangle \eta},\\
\mathcal{T}^{^{\,\mbox{\tiny{3rd}}}}_7 &= \Theta \sigma_\eta^{~\langle \mu}\Omega^{\nu\rangle \eta},\quad &
\mathcal{T}^{^{\,\mbox{\tiny{3rd}}}}_8 &= \Theta \Omega_\eta^{~\langle \mu}\Omega^{\nu\rangle \eta},
\quad & \mathcal{T}^{^{\,\mbox{\tiny{3rd}}}}_9 &=\sigma_\eta^{~\langle \mu}\sigma^{\nu\rangle \alpha}
\sigma_\alpha^{~\eta},\\
\mathcal{T}^{^{\,\mbox{\tiny{3rd}}}}_{10} &= \Omega_\eta^{~\langle \mu}
\Omega^{\nu\rangle \alpha} \sigma_\alpha^{~\eta}, \quad & \mathcal{T}^{^{\,\mbox{\tiny{3rd}}}}_{11}
&= \sigma_\eta^{~\langle \mu}\Omega^{\nu\rangle \alpha} \Omega_\alpha^{~\eta},
\quad & \mathcal{T}^{^{\,\mbox{\tiny{3rd}}}}_{12} &= \Omega_\eta^{~\langle\mu}
\Omega^{\nu\rangle \alpha} \Omega_\alpha^{~\eta},\\
\mathcal{T}^{^{\,\mbox{\tiny{3rd}}}}_{13} &= \nabla_\perp^2 \sigma^{\expva{\mu\nu}},\quad & 
\mathcal{T}^{^{\,\mbox{\tiny{3rd}}}}_{14} &= \nabla_{_\perp}^{~\langle \mu} \nabla_{_\perp}^{~\nu \rangle}\Theta,
\quad & \mathcal{T}^{^{\,\mbox{\tiny{3rd}}}}_{15} &=\Theta\nabla_{_\perp}^{~\langle \mu}
\nabla_{_\perp}^{~\nu \rangle}\ln s,\\
\mathcal{T}^{^{\,\mbox{\tiny{3rd}}}}_{16} &= \Theta \nabla_{_\perp}^{~\langle \mu}\ln s
\nabla_{_\perp}^{~\nu \rangle}\ln s, \quad & \mathcal{T}^{^{\,3rd}}_{17} &= \sigma_\alpha^{~\langle \mu}
\nabla_{_\perp}^{~\nu \rangle} \nabla_{_\perp}^\alpha \ln s, \quad & \mathcal{T}^{^{\,3rd}}_{18}
&=\Omega_\alpha^{~\langle \mu} \nabla_{_\perp}^{~\nu \rangle} \nabla_{_\perp}^\alpha \ln s,\\
\mathcal{T}^{^{\,\mbox{\tiny{3rd}}}}_{19} &= \sigma_\alpha^{~\langle \mu} \nabla_{_\perp}^{~\nu \rangle}\ln s
\nabla_{_\perp}^\alpha \ln s, \quad & \mathcal{T}^{^{\,\mbox{\tiny{3rd}}}}_{20} &= \Omega_\alpha^{~\langle \mu}
\nabla_{_\perp}^{~\nu \rangle}\ln s \nabla_{_\perp}^\alpha \ln s, \quad &
\mathcal{T}^{^{\,\mbox{\tiny{3rd}}}}_{21} &=\nabla_{_\perp}^\alpha\sigma^{\expva{\mu\nu}}
\nabla_{_\perp\alpha} \ln s,\\
\mathcal{T}^{^{\,\mbox{\tiny{3rd}}}}_{22} &= \nabla_{_\perp}^\alpha\Omega^{\expva{\mu\nu}}
\nabla_{_\perp\alpha} \ln s,\quad &
\mathcal{T}^{^{\,\mbox{\tiny{3rd}}}}_{23} &= \nabla_{_\perp}^{~\langle \mu}\Theta
\nabla_{_\perp}^{~\nu \rangle} \ln s,\quad &
\mathcal{T}^{^{\,\mbox{\tiny{3rd}}}}_{24} &= \nabla_{_\perp}^\alpha\sigma_\alpha^{~\langle \mu} \nabla_{_\perp}^{~\nu \rangle} \ln s,\\
\mathcal{T}^{^{\,3rd}}_{25} &= R\sigma^{\mu\nu}, \quad &\mathcal{T}^{^{\,3rd}}_{26} &= u^\alpha u^\beta R_{\alpha \beta}\sigma^{\mu\nu},
\quad & \mathcal{T}^{^{\,\mbox{\tiny{3rd}}}}_{27} &= u^\gamma \nabla^\alpha R_{\gamma~~~~\alpha}^{~\expva{\mu\nu}},\\
\mathcal{T}^{^{\,\mbox{\tiny{3rd}}}}_{28} &= u^\gamma u^\alpha D R_{\gamma~~~~\alpha}^{~\expva{\mu\nu}},\quad & 
\mathcal{T}^{^{\,\mbox{\tiny{3rd}}}}_{29} &= DR^{\expva{\mu\nu}},\quad & 
\mathcal{T}^{^{\,\mbox{\tiny{3rd}}}}_{30} &= u^\gamma \nabla^{\langle \mu}R^{\nu\rangle}_{~~\gamma},\\
\mathcal{T}^{^{\,\mbox{\tiny{3rd}}}}_{31} &= u^\gamma \left(\nabla_{_\perp}^\alpha\ln s\right) R_{\gamma~~~~\alpha}^{~\expva{\mu\nu}},\quad & 
\mathcal{T}^{^{\,\mbox{\tiny{3rd}}}}_{32} &= \sigma^{\gamma \alpha} R_{\gamma~~~~\alpha}^{~\expva{\mu\nu}},\quad & 
\mathcal{T}^{^{\,\mbox{\tiny{3rd}}}}_{33} &= u^\gamma u^\alpha \sigma^{\beta \langle \mu} R^{\nu\rangle}_{~~\gamma\alpha\beta},\\
\mathcal{T}^{^{\,\mbox{\tiny{3rd}}}}_{34} &= u^\gamma u^\alpha \Omega^{\beta \langle \mu} R^{\nu\rangle}_{~~\gamma\alpha\beta},
\quad & \mathcal{T}^{^{\,\mbox{\tiny{3rd}}}}_{35} &= \Theta u^\gamma u^\alpha  R_{\gamma~~~~\alpha}^{~\expva{\mu\nu}},\quad &  
\mathcal{T}^{^{\,\mbox{\tiny{3rd}}}}_{36} &= u^\gamma R_\gamma^{\langle \mu}\nabla_{_\perp}^{\nu\rangle}\ln s,\\
\mathcal{T}^{^{\,\mbox{\tiny{3rd}}}}_{37}  &= \Theta R^{\expva{\mu\nu}},\quad &
\mathcal{T}^{^{\,\mbox{\tiny{3rd}}}}_{38} &=\sigma_\gamma^{~\langle \mu } R^{\nu\rangle \gamma},\quad & 
\mathcal{T}^{^{\,\mbox{\tiny{3rd}}}}_{39} &=\Omega_\gamma^{~\langle \mu } R^{\nu\rangle \gamma}.
\end{aligned}
\end{equation}

This list of independent third-order tensor structures is reduced by 6
structures when compared to the previous tensors obtained in Ref. \cite{Grozdanov:2015kqa}.
The reason for such a difference is the same as we  pointed out previously for third-order scalars.
To be explicit, the structures in each one of the sets of tensors
$\{\nabla_{\perp \alpha}\nabla_{\perp}^\alpha\nabla_{\perp}^{\langle \mu} u^{\nu\rangle}$,$\nabla_{\perp \alpha}
\nabla_{\perp}^{\langle \mu} \nabla_{\perp}^\alpha u^{\nu\rangle}$,$\nabla_{\perp}^{\langle \mu}
\nabla_{\perp \alpha}\nabla_{\perp}^\alpha u^{\nu\rangle}\}$, $\{\nabla_{\perp \alpha}
\nabla_{\perp}^{\langle \mu}\nabla_{\perp}^{\nu\rangle} u^\alpha$,$\nabla_{\perp}^{\langle \mu}
\nabla_{\perp\alpha}\nabla_{\perp}^{\nu\rangle} u^\alpha$,$\nabla_{\perp}^{\langle\mu}
\nabla_{\perp}^{\nu\rangle} \nabla_{\perp \alpha}u^\alpha\}$, $\{\nabla_{\perp}^\alpha\ln s
\nabla_{\perp \alpha}\nabla_{\perp}^{\langle\mu} u^{\nu\rangle}$, $\nabla_{\perp}^\alpha\ln s
\nabla_{\perp}^{\langle\mu}\nabla_{\perp\alpha} u^{\nu\rangle}\}$, $\{\nabla_{\perp}^{\langle\mu}\ln s
\nabla_{\perp\alpha}\nabla_{\perp }^{\nu\rangle} u^\alpha$,$\nabla_{\perp}^{\langle\mu}\ln s \nabla_{\perp}^{\nu\rangle}\nabla_{\perp\alpha}u^\alpha \}$
differ only by the ordering of covariant derivatives and hence six of them are redundant
in the gradient and can be removed from the list. 

Therefore, the general third-order corrections to the energy-momentum tensor
come with $58=19+39$ transport coefficients, that we label as $\chi_a$ for the scalars
and as $\gamma_b$ for tensors, and the third-order contribution to the
energy-momentum tensor may be represented in the form
\begin{equation}
\Pi_{\scriptscriptstyle{(3)}}^{\mu\nu} =\sum_{a=1}^{19} \chi_a\, \mathcal{S}^{^{\mbox{\tiny{3rd}}}}_a\Delta^{\mu\nu}
+\sum_{b=1}^{39} \gamma_b  \left(\mathcal{T}^{^{\,\mbox{\tiny{3rd}}}}_b\right)^{\mu\nu}.  
\end{equation}

The reduction of the number of scalar and tensor corrections means a simplification of the
higher-order hydrodynamics that is very welcome. The results presented here provide a
reduction of $10$ transport coefficients from the previous description of the third-order
hydrodynamics for a non-conformal system.

\section{Relativistic hydrodynamics of conformal fluids}
\label{Sec:ConformalFluids}

\subsection{Conformal covariance} 
 
Conformal symmetry is associated with the light-like particles in relativistic dynamics, it is the light-cone symmetry. A fluid build up from massive particles can approach conformalsymmetry when kinetic energy is very large in comparison to  other energy content of each fluid component.
Such a situation is realized, for instance, in QGP formed in heavy ion collision.
Now the invariance of the hydrodynamic system under conformal, or Weyl, scaling should be taken into account in order to build up correctly the respective gradient expansion.\footnote{Here we treat Weyl symmetry and conformal symmetry on the same foot even though there are subtle differences between them. A discussion on this subject can be found in Refs. \cite{Jackiw:2005su,Jackiw:2011vz,Farnsworth:2017tbz}.} 
The energy-momentum tensor $T^{\mu\nu}$ of a system that is invariant under a Weyl scaling $g_{\mu\nu}\to e^{2\phi(x)}g_{\mu\nu}$  should transform as the tensor density, $T^{\mu\nu}\to e^{-(d-2)\phi(x)}T^{\mu\nu}.$
Moreover, the energy-momentum tensor of a conformal theory should be traceless \cite{Baier:2007ix}. This implies that scalar corrections to the pressure of a conformal fluid vanishes, eliminating all transport coefficients coming from higher-order scalars.
Since conformal scalars are absent in the gradient expansion we consider such structures separately in Appendix \ref{Sec:AppendixB}.

The program of finding the hydrodynamic expansion for conformal fluids at third order is considered in \cite{Grozdanov:2015kqa}
where the authors establish an approach to build admissible structures for the energy-momentum tensor.
Despite their approach presents some positive features,
problems arise in a higher-order expansion. 
The strategy adopted by the authors to find the higher-order structures is to take the list of all
non-conformal structures, considering 
for each one its behaviour under conformal scaling, and then finding linear combinations in such a way that inhomogeneous terms are eliminated and 
the final structure is conformally covariant. The strategy works fine at first and second orders but the work becomes very hard at third order.
For third-order gradients there is an accumulation of inhomogeneous terms when they are submitted to a
Weyl scaling and to find the 
linear combinations that eliminate all of such terms is not an easy task. 
Alternatively, here we  build the gradient expansion by using structures that are, by construction,
covariant under symmetry transformations.  In the same way as  
the partial derivative is corrected by introducing the Christofell
connection terms to obtain covariant structures under diffeomorphism transformations, 
the covariant derivative needs to be corrected  
to obtain gradients that are covariant under Weyl scaling too. Namely,  
we define the Weyl-covariant derivative acting in the fundamental fields in such a way to preserve
their character of tensor density under Weyl scaling.

In the context of relativistic hydrodynamics the Weyl-covariant derivative was introduced
in \cite{Loganayagam:2008is} using a vector connection.  It was recently shown in \cite{Diles:2017azi}
that the formulation of the Weyl covariance by a minimal coupling mechanism is in fact well defined.
In such an approach the scaling factor of a fundamental field plays the role of the gauge charge associated with the symmetry under local Weyl scaling. The minimal coupling for Weyl symmetry works as
follows. Let $\psi$ be a tensor density of weight $\omega$, we minimally couple it with Weyl connection $\mathcal{A}_\mu$ by replacing $\partial_\mu \psi$ by $(\partial_\mu + \omega \mathcal{A}_\mu)\psi$
so that under conformal scaling it transforms as $(\partial_\mu + \omega \mathcal{A}_\mu)\psi\to
e^{-\omega \phi}(\partial_\mu + \omega \mathcal{A}_\mu)\psi$. The minimal coupling prescription is well defined for any composition of derivatives and is a direct mechanism to construct high order
gradients that are Weyl covariant. To implement Weyl covariance in this way we need a vector field $\mathcal{A}_\mu$ that transforms as a connection under Weyl scaling, i.e., $ \mathcal{A}_\mu\to \mathcal{A}_\mu+\partial_\mu\phi$.
It was noted in \cite{Loganayagam:2008is} that, in hydrodynamics, the combination
of first-order gradients  
\begin{equation}
\mathcal{A}_\mu = -\frac{1}{d-1}(\nabla_{_\perp \mu} \ln s + \Theta u_\mu)  \label{connection}
\end{equation}
transforms in the mentioned form, i.e., transforms as a connection under Weyl scaling.
The presence of a gauge connection for the Weyl symmetry is a necessary and sufficient condition to
ensure that the minimal coupling prescription fulfills Weyl covariance. In order to build up
the gradient expansion, we use only Weyl-covariant derivatives of the fundamental fields. 
 These relations are discussed in detail in \cite{Loganayagam:2008is},
and many equivalences  
obtained in that work are used here to rule out redundant corrections in the
energy-momentum tensor of the conformal fluid. In the following the Weyl-covariant derivative
is denoted by $\mathcal{D}_\mu$.

It is remarkable that the shear and vorticity tensors are naturally Weyl-covariant objects and can be defined
as $\sigma^{\mu\nu} = \mathcal{D}^\mu u^\nu+\mathcal{D}^\nu u^\mu$ and
$\Omega^{\mu\nu} = \mathcal{D}^\mu u^\nu-\mathcal{D}^\nu u^\mu$.  However,
neither the expansion $\Theta$ nor the entropy gradient $\nabla_{_\perp} \ln s$ are Weyl covariant structures, but 
they combine to form the gauge connection shown in eq. (\ref{connection}).
A connection is in fact not covariant, but it gives rise to a covariant field tensor $\mathcal{F}$
defined by its exterior derivative, 
\begin{equation}
\mathcal{F}=d\mathcal{A}=\mathcal{F}_{\mu\nu}dx^\mu\wedge dx^\nu,\qquad
\mathcal{F}_{\mu\nu}=\partial_\mu \mathcal{A}_\nu-\partial_\nu \mathcal{A}_\mu.
\end{equation}

The usual curvature tensors related to the Riemann tensor are not invariant under conformal transformations
and, as pointed out in \cite{Diles:2017azi}, there are two ways to deal with this situation. One can take combinations
of the Riemann tensor and its contractions by building the  Weyl tensor $C^\mu_{~\nu\alpha\beta}$, or
one can couple the metric to the Weyl connection through the minimal coupling prescription by obtaining
the conformal Riemann tensor $\mathcal{R}^\mu_{~\nu\alpha\beta}$\footnote{The Weyl tensor can be obtained as a combination of the conformal Riemann
($\mathcal{R}$) and the Weyl curvature ($\mathcal{F}$), 
$\displaystyle{ C_{\mu \nu\eta\sigma} = \mathcal{R}_{\mu \nu\eta\sigma} +
\frac{1}{d-2}\delta^\alpha_{~[\mu}g_{\nu][\eta}\delta_{\sigma]}^{~\lambda}\left(\mathcal{R}_{\alpha\lambda} - \frac{1}{2(d-1)}\mathcal{R}g_{\alpha\lambda}\right) +
\mathcal{F}_{\mu \nu}g_{\eta\sigma}. }$}.
Here we take the later prescription in such a way that our convention agrees with
Ref.~\cite{Loganayagam:2008is}, and then we may directly use the identities presented in that work. 
Therefore, the set of irreducible gradients we have at hand to define the gradient expansion for a conformal fluid is $\{\sigma^{\mu\nu},\,\Omega^{\mu\nu},\,\mathcal{F}^{\mu\nu},\,\mathcal{R}_{\mu \nu\eta\sigma}\}.$  
Note that $\mathcal{F}$ is anti-symmetric and second order in derivatives. Due to its anti-symmetry it will not appear in the gradient expansion at second order and it will appear at first time
in the third-order corrections for the conformal fluid. Two first-order terms
$\{\Theta,\,\nabla_{_\perp}^\mu\ln s\}$ allowed in the non-conformal case are now forbidden.
These two structures together are directly responsible for $1+0$ transport coefficients at first order,
$2+5$ at second order and $12+21$ at third order. This is the main reason why the gradient expansion
of a conformal fluid is strongly simplified in comparison with the case of a
non-conformal fluid.

\subsection{The first- and second-order conformal hydrodynamics}
\label{sec:order1and2}

Conformal symmetry requires the energy-momentum tensor to be traceless so that the pressure of the conformal fluid is constrained to be proportional to the energy density, $p(\varepsilon) = \varepsilon/(d-1)$
and the scalar corrections vanish $\left(\mathcal{P}_{\scriptscriptstyle{(i)}}=0\right)$ in the Landau frame.
Hence the ideal energy-momentum tensor of a conformal fluid may be cast into the form
\begin{equation}
T^{\mu\nu}_{\mbox{\scriptsize{ideal}}} =   \varepsilon u^\mu u^\nu + \frac{\varepsilon}{d-1}\Delta^{\mu\nu}.
\end{equation}

Conformal symmetry also changes the way in which the equations of motion are
written in terms of the energy-momentum tensor, since the usual covariant derivative $\nabla_\mu$ must be replaced by the Weyl-covariant derivative ${\cal D}_\mu$ mentioned in the last section.
The equations of motion of the conformal hydrodynamics are then $\mathcal{D}_\mu T^{\mu\nu}=0$.
In view of the form of the Weyl connection (\ref{connection}), it follows that $\mathcal{D}_\mu u^\mu =0$, 
and the zeroth-order equations of motion for conformal fluids give us $\mathcal{D}^\mu \varepsilon =0.$
 
As a consequence, all the higher-order energy gradients also vanish and can be excluded from the gradient expansion for a conformal fluid \cite{Loganayagam:2008is}.
At first order, the gradient expansion becomes trivial since there is only one first-order structure allowed, $\sigma^{\mu\nu}$,  with one associated transport coefficient, $\eta$, and so at this order the correction for the energy-momentum tensor is just
\begin{equation}
\Pi_{\scriptscriptstyle{(1)}}^{\mu\nu}= -\eta\sigma^{\mu\nu}.
\end{equation}

At second order there are more possibilities, but not so much as in comparison with the non-conformal case. 
To build the second-order tensors we can take symmetric derivatives of first-order structures, in particular
we can take longitudinal derivatives $u^\mu\mathcal{D}_\mu$, as well as symmetric products of first-order structures.
Accordingly, the set of allowed structures is given by
$\{{\sigma_{\alpha}}^{\langle \mu}\,\sigma^{\nu\rangle \alpha},\,{\sigma_{\alpha}}^{\langle \mu}\,
\Omega^{\nu\rangle \alpha},\,
\Omega_{\alpha}^{\;\,\langle \mu}\,\Omega^{\nu\rangle \alpha},\, u^\alpha\mathcal{D}_\alpha \sigma^{\expva{\mu\nu}},
\, u^\alpha\mathcal{D}_\alpha \Omega^{\expva{\mu\nu}},\, u^\alpha u^\beta
\mathcal{R}_{\alpha~~~~\beta}^{~\expva{\mu\nu}},\,\mathcal{R}^{\expva{\mu\nu}}\} $.
Notice, however, that not all of these elements are relevant. In fact, due to the anti-symmetry of the
vorticity $\Omega^{\mu\nu},$ we have identically that $u^\alpha\mathcal{D}_\alpha \Omega^{\expva{\mu\nu}}=0$.
 
Algebraic identities allow us to express $u^\alpha u^\beta\mathcal{R}_{\alpha~~~~\beta}^{~\expva{\mu\nu}}$
as a linear combination of $\sigma_{\alpha}^{\,\,\langle \mu}\sigma^{\nu\rangle\alpha}$,
${\Omega_{\alpha}}^{\langle \mu}\,\Omega^{\nu\rangle \alpha}$ and $\mathcal{R}^{\expva{\mu\nu}}$.
Using such relations we obtain only five independent second-order tensors,
\begin{equation}
\mathfrak{T}^{^{\mbox{\tiny{2nd}}}}_1=\sigma_\alpha^{~\langle \mu}\sigma^{\nu\rangle \alpha},\;\;
\mathfrak{T}^{^{\mbox{\tiny{2nd}}}}_2=\Omega_\alpha^{~\langle \mu}\Omega^{\nu\rangle \alpha} ,\;\;
\mathfrak{T}^{^{\mbox{\tiny{2nd}}}}_3=\sigma_\alpha^{~\langle \mu}\Omega^{\nu\rangle \alpha},\;\;
\mathfrak{T}^{^{\mbox{\tiny{2nd}}}}_4=u^\alpha\mathcal{D}_\alpha \sigma^{\expva{\mu\nu}},\;\;
\mathfrak{T}^{^{\mbox{\tiny{2nd}}}}_5=\mathcal{R}^{\expva{\mu\nu}},
\end{equation}
where again we have omitted the tensor indexes on the left hand side of the equations.
At last, using the above notation, the second-order corrections to the energy-momentum tensor of a conformal fluid may be written as
\begin{equation}
\Pi_{\scriptscriptstyle{(2)}}^{\mu\nu}= \sum_{a=1}^{5} \upsilon_a
\left(\mathfrak{T}^{^{\mbox{\tiny{2nd}}}}_a\right)^{\mu\nu},  
\end{equation}
where the necessity of only five transport coefficients is apparent. 
 
The last result showing only five second-order independent correction terms for a conformal fluid,
requiring only $5$ additional transport coefficients, is in complete agreement with the present literature.
The manifestation of conformal symmetry indicates that the 
speeds involved are close to the speed of light,
so that relativistic effects are dominant and we must take seriously the causality issue.
We know that  truncating 
the hydrodynamic expansion 
at first order 
one faces the problem of instantaneous propagation of sound \cite{Romatschke:2009im}.
Therefore, one needs to go at least up to second order in gradient expansion to find a
consistent description of conformal fluids, and this requires in general the introduction
of $6$ transport coefficients, one at first order and five at second order.

\subsection{The third-order conformal hydrodynamics}
\label{Sec:ThirdHydro}

As commented above for the first- and second-order conformal hydrodynamics, the relevant independent corrections to $T^{\mu\nu}_{\mbox{\scriptsize{ideal}}}$ at third order in the gradient expansion are also two-rank tensorial structures.
However, for a question of completeness, the sets of
independent third-order conformal scalars and vectors, which do not appear
in previous works, are presented in Appendix \ref{Sec:AppendixB}.

In order to find the third-order tensors we can perform symmetric products of first- and second-order gradients, and take symmetric derivatives of second-order structures.
The third-order corrections for conformal fluids may be obtained
in the same way as the third-order corrections for the non-conformal fluids,
the difference between the two cases is the set of building blocks - covariant gradients -
and the algebraic identities they satisfy in each case.

Considering the transversality of $\sigma^{\mu\nu}$ and $\Omega^{\mu\nu}$, one can show that
$u_\alpha\mathcal{D}_\beta\left(\sigma^{\beta\langle \mu}\sigma^{\nu\rangle \alpha}\right) = -\frac{1}{2} \sigma_{\alpha\beta}\sigma^{\alpha\langle \mu}\sigma^{\nu\rangle \beta}$ and    
$u_\alpha\mathcal{D}_\beta\left(\Omega^{\beta\langle \mu}\Omega^{\nu\rangle \alpha}\right)=
-\Omega_{\alpha\beta}\Omega^{\alpha\langle \mu}\Omega^{\nu\rangle \beta}$. 
Now the tensor $\Omega_{\alpha\beta}\,\Omega^{\alpha\langle \mu}\,\Omega^{\nu\rangle \beta}$
vanishes identically due to the 
anti-symmetric property of the vorticity. 
We can also remove from the list one tensor out of 
$\mathcal{D}_\alpha\mathcal{D}^{\langle\mu}\sigma^{\nu\rangle\alpha}$,
$\mathcal{D}_\alpha\mathcal{D}^{\langle\mu}\Omega^{\nu\rangle\alpha}$
and $\Delta^{\alpha \beta}\mathcal{D}_\alpha\mathcal{D}_\beta\sigma^{\expva{\mu\nu}}$,
since the sum of the first two differs from the latter 
essentially by the order of the derivatives.
The tensor $\Omega^{\alpha\beta}\,\mathcal{R}^{~\expva{\mu\nu}}_{\alpha~~~~\beta}$ appears as
a candidate for the list of independent structures but it also vanishes identically.
We then find  only 19 completely independent third-order conformally covariant tensors, which are given by
 
\begin{equation}
\begin{aligned}[7]
\mathfrak{T}^{^{\mbox{\tiny{3rd}}}}_1 &= \Omega^2\sigma^{\mu\nu},&
\mathfrak{T}^{^{\mbox{\tiny{3rd}}}}_6 &=\mathcal{D}_\alpha\mathcal{D}^{\langle\mu}\sigma^{\nu\rangle\alpha},&
\mathfrak{T}^{^{\mbox{\tiny{3rd}}}}_{11} &= u_\alpha\mathcal{D}^{\langle\mu}\mathcal{F}^{\nu\rangle\alpha},&
\mathfrak{T}^{^{\mbox{\tiny{3rd}}}}_{16} &=\mathcal{R}_\alpha^{~\langle \mu}\Omega^{\nu\rangle \alpha},\\
\mathfrak{T}^{^{\mbox{\tiny{3rd}}}}_2 & =\sigma^2\sigma^{\mu\nu},&
\mathfrak{T}^{^{\mbox{\tiny{3rd}}}}_7 &= \Delta^{\alpha\beta}\mathcal{D}_\alpha\mathcal{D}_\beta\sigma^{\expva{\mu\nu}},&
\mathfrak{T}^{^{\mbox{\tiny{3rd}}}}_{12} &=\mathcal{F}_\alpha^{~\langle \mu}\sigma^{\nu\rangle \alpha},&
\mathfrak{T}^{^{\mbox{\tiny{3rd}}}}_{17} &=\mathcal{D}^{\langle \mu} \mathcal{R}^{\nu\rangle}_{~~\alpha}u^\alpha,\\
\mathfrak{T}^{^{\mbox{\tiny{3rd}}}}_3 &=\sigma_{\alpha\beta}\sigma^{\alpha\langle \mu}\sigma^{\nu\rangle \beta},&
\mathfrak{T}^{^{\mbox{\tiny{3rd}}}}_8 &=u^\alpha\mathcal{D}_\alpha \sigma^{\beta\langle \mu}\sigma^{\nu\rangle}_{~~\beta},&
\mathfrak{T}^{^{\mbox{\tiny{3rd}}}}_{13}&=\mathcal{F}_\alpha^{~\langle \mu}\Omega^{\nu\rangle \alpha},&
\mathfrak{T}^{^{\mbox{\tiny{3rd}}}}_{18} &=u^\alpha\mathcal{D}^\beta\mathcal{R}^{~\expva{\mu\nu}}_{\alpha~~~~\beta},\\
\mathfrak{T}^{^{\mbox{\tiny{3rd}}}}_4 &= \sigma_{\alpha\beta}\Omega^{\alpha\langle \mu}\Omega^{\nu\rangle \beta},&
\mathfrak{T}^{^{\mbox{\tiny{3rd}}}}_9 &=u^\alpha\mathcal{D}_\alpha \Omega^{\beta\langle \mu}\sigma^{\nu\rangle}_{~~\beta},&
\mathfrak{T}^{^{\mbox{\tiny{3rd}}}}_{14} &=\mathcal{R}\sigma^{\mu\nu},&
\mathfrak{T}^{^{\mbox{\tiny{3rd}}}}_{19} &= \sigma^{\alpha\beta}\mathcal{R}^{~\expva{\mu\nu}}_{\alpha~~~~\beta}.\\
\mathfrak{T}^{^{\mbox{\tiny{3rd}}}}_5 &=\Omega_{\alpha\beta}\sigma^{\alpha\langle \mu}\Omega^{\nu\rangle \beta},&
\mathfrak{T}^{^{\mbox{\tiny{3rd}}}}_{10}
&=u^\alpha\mathcal{D}_\alpha \Omega^{\beta\langle \mu}\Omega^{\nu\rangle}_{~~\beta},&
\mathfrak{T}^{^{\mbox{\tiny{3rd}}}}_{15} &= \mathcal{R}_\alpha^{~\langle \mu}\sigma^{\nu\rangle \alpha}.&
\end{aligned}
\end{equation}
 
By labeling the third-order transport coefficients as $\vartheta_b$, we finally obtain the
complete third-order correction to the energy-momentum tensor of a conformal fluid in a closed form,
 
\begin{equation}
\Pi_{\scriptscriptstyle{(3)}}^{\mu\nu}=\sum_{b=1}^{19}\vartheta_b \left(\mathfrak{T}^{^{\mbox{\tiny{3rd}}}}_b\right)^{\mu\nu}.  
\end{equation}
 
Considering that the generalization of the Navier-Stokes equation for a conformal fluid is given by
the conformal(Weyl)-covariant divergence of its energy-momentum tensor, 
we then obtain the equations of motion for a conformal fluid up to third order in the gradient  expansion by
\begin{equation}
\mathcal{D}_\mu\left(T^{\mu\nu}_{\mbox{\scriptsize{ideal}}}+\Pi_{\scriptscriptstyle{(1)}}^{\mu\nu}
+\Pi_{\scriptscriptstyle{(2)}}^{\mu\nu}+\Pi_{\scriptscriptstyle{(3)}}^{\mu\nu}\right)= 0.
\end{equation} 

The list of independent third-order tensor perturbations, that is found here to be 19,
fixes the number of additional transport coefficients in conformal hydrodynamics. 
It is a dramatic reduction in comparison with the non-conformal hydrodynamics,
where $58$ third-order transport coefficients are needed, and so the constraints
coming from the conformal symmetry eliminate $39$ of them. We have found here only
19 independent conformal tensors and it differs from the previous results of Ref. \cite{Grozdanov:2015kqa}
which presents 20 tensors. The reduction of such a number by unity shows that there exists a non-trivial linear combination of the 20 previously known tensors that vanishes identically.

We call  attention to the necessity of minimizing the redundancies when dealing with higher-order terms, which is a difficult task since the number of possibilities grows up exponentially for a higher-order expansion. 
In the particular case of conformal symmetry, the use of conformally covariant derivatives is essentially necessary, in the very same way one needs to implement the usual covariant derivative
when considering a fluid living in a curved space-time. 
The way we proceed here differs from what was previously done in order to find the third-order tensors for conformal fluids in  \cite{Grozdanov:2015kqa} exactly on this issue. There the authors first list all non-conformal structures and then look for the linear combinations that transform covariantly under Weyl scaling.  In this sense the results we present in this and in the previous sections do not contradict the
results of \cite{Grozdanov:2015kqa}, but rather complement them by improving the systematic methodology introduced there, by reducing the number of transport coefficients at third order in non-conformal and conformal hydrodynamics, and by allowing us to obtain a complete list of independent conformal structures up to third order.

\subsection{Comments on phenomenology}

Even though hydrodynamics may be seen as an effective description of the collective effects of fundamental interactions, it does not deal with the interactions at all, in practice it deals with the symmetry of the interactions we expect to manifest in the macroscopic dynamics. We build up a gradient expansion for the energy-momentum tensor including all possible corrections for each order of derivatives and each term that enter as a correction comes with a free parameter,
the transport coefficient. The equations of motion and other algebraic relations reveal linear dependencies
between the possible corrections. When this is the  case we can write one correction as a linear combination of the others  and the associated transport coefficient is eliminated 
from the expansion. 
However, this does not mean that the transport coefficient vanishes; the list of independent transport coefficients is reduced by appropriately defining a
new transport coefficient as a linear combination of the old ones. 
For example, in Refs.  \cite{Jaiswal:2013vta, Jaiswal:2014raa, Chattopadhyay:2014lya}, the kinetic theory is used to compute the equations
of motion for a conformal fluid including the contributions of third-order gradients and 14 different third-order tensors appear in the equations of motion, although there are only two transport coefficients ($\tau_\pi, \beta_\pi$). We find here 13 transport coefficients associated with the set of independent third-order conformal tensors in flat space that are completely arbitrary. Theoretical consistence of both approaches is ensured if one can reproduce the results of the kinetic theory for some particular choice of transport coefficients. We expect that it can be done, but it is not a simple task and will not be developed here.

On the other hand, the absence of scalar perturbations in the energy-momentum tensor of a conformal fluid (due to the vanishing of its trace)  has a very direct consequence. The non-vanishing of any  transport
coefficient associated with a scalar correction at any order of the gradient expansion implies that the system is not conformally invariant. Therefore, the presence of a non-vanishing  scalar at a given order of the gradient expansion is enough to report the breaking of conformal symmetry at that order. It means that even the fluid looking conformal, for instance, at second-order hydrodynamics, it can be verified to be non-conformal when submitted to a third-order analysis.

\section{Dispersion relations}
\label{Sec:DispersionRelations}

\subsection{The general approach}

The dispersion relations of waves propagating in a fluid are obtained 
by considering the equations of motion for small perturbations in the momentum space.
Such equations may be found by applying the standard techniques of linear perturbation
theory to the set of hydrodynamic equations for the fluid, which are
obtained from the gradient expansion developed in the previous sections. At a given order $n$ in gradient expansion,
the equations of motion for the fluid read $\nabla_\mu T_{\scriptscriptstyle{(n)}}^{\mu\nu} = 0$, 
where $T_{\scriptscriptstyle{(n)}}^{\mu\nu}$ is the energy-momentum tensor containing up to $n$ derivatives.
This is a vector equation and can be locally projected along the flow $u_\nu$ as $u_\nu\nabla_\mu
T_{\scriptscriptstyle{(n)}}^{\mu\nu} = 0,$ as well as perpendicularly to the flow as
$\Delta^{\alpha}_{~\nu}\nabla_\mu T_{\scriptscriptstyle{(n)}}^{\mu\nu}=0$.
The perturbations are then small deviations from a fixed configuration that corresponds
to a solution of the equations of motion. 
Here we assume that spacetime is flat 
and analyze perturbations of a static configuration, 
where both the velocity and the entropy are constant throughout the system,  $u^\mu(x)=u_0^\mu$
and $\ln s(x)=\ln s_0$.  In the perturbed configuration the degrees of freedom read
\begin{equation}\label{perturb1}
u^\mu(x) =  u_0^\mu + \delta u^\mu(x),\qquad\quad \ln s(x) = \ln s_0 + \delta \ln s(x).
\end{equation}
where the $\delta$ symbol indicates small quantities.
We now insert the expressions shown in Eq.~\eqref{perturb1} into the equations of motion for the fluid and look
for the linear terms in $\delta u$ and $\delta\ln s$. 
The resulting linearized equations describe the propagation of waves in the fluid. 

To work in the momentum space, we Fourier transform the fluctuations
\begin{equation} 
\delta u(x) = \int d^{d}x\,e^{ik_\mu x^\mu}\delta u(k),\qquad
\delta \ln s(x) = \int d^{d}x\,e^{ik_\mu x^\mu}\delta \ln s(k)\,.
\end{equation}
As known from the linear perturbation theory in flat spacetime, the set of differential equations
in the configuration space translates into a set of algebraic equations for the fields
$\delta u(k)$, $\delta \ln s(k)$ in the corresponding Fourier space, since the spacetime
derivatives are replaced by the wavevector as $\partial_\mu\to ik_\mu$. The solutions of such
equations furnish the dispersion relations. Notice that a term of the energy-momentum tensor
with $n$ spacetime derivatives, or with spacetime derivatives of order $n$,  results in a power
$k^{n+1}$ in the (algebraic) equations of motion for the perturbations in the Fourier space.
Hence a $n$-th order hydrodynamics provides an approximation up to power $n+1$ in the dispersion relations.

The perturbation equations may be conveniently projected along the flow and perpendicularly to the flow,
since each one of these projections plays a different role. So we introduce here the
wavevector projections  
\begin{equation} \label{kprojections}
    k_{_\parallel}= k_\mu u^\mu, \qquad k_{_\perp}^\mu =\Delta^\mu_{~\nu}k^\nu,
\end{equation}
respectively, along the flow and perpendicular to the flow.

The parallel projection of the equations of motion reads
\begin{equation}
u^\mu\nabla_\mu \ln s + \Theta + (\nabla_\mu u_\nu) \Pi_{\scriptscriptstyle{(n)}}^{\mu\nu}= 0\,,
\label{parallel_equation}
\end{equation}
but linearization rules out the last term. In fact, 
the products and powers of derivatives, like $\partial u\, \partial \ln s$, $(\partial u)^2$, etc.,
are higher order in the amplitudes and so are ruled out by linearization,
which selects only the terms with successive derivatives of the same quantity, such as $\partial^2u, \partial^2\ln s.$
Hence, independently of the gradient-expansion order, the equation of motion \eqref{parallel_equation}
reduces to
\begin{equation}
 u^\mu\partial_\mu(\delta \ln s) + \Delta^\mu_{~\nu}(\partial_\mu\delta u^\nu) = 0.
\end{equation} 
In terms of the velocity perturbations, the last differential equation gives the
perturbation in the entropy,
 \begin{equation}
\delta \ln s = -\frac{1}{k_{_\parallel}}k_{_\perp}^\mu \delta u_{\mu},
\label{longitf}
\end{equation}
so that the parallel equation of motion can be used to eliminate $\delta \ln s$ in terms of $ \delta u_\mu$.

The perpendicular projection of the equations of motion is strongly dependent on the order
of the gradient expansion. Its linearization yields 
\begin{equation}
u^\nu\partial_\nu \left(\delta u^\mu\right) + c_s^2\Delta^{\mu\nu}\partial_\nu
\left(\delta\ln s\right) + \frac{1}{\epsilon + p} \Delta^\mu_{~\nu}
\partial_\alpha \left(\sum_{i=1}^n \delta\Pi_{\scriptscriptstyle{(i)}}^{\nu\alpha}
\right) = 0,
\label{dispersion}
\end{equation}
where $\delta\Pi_{\scriptscriptstyle{(i)}}^{\mu\nu}$ is the linear part of the energy-momentum tensor correction at order $i$.
In order to find out the dispersion relations, we need to express $\partial_\alpha \delta\Pi_{\scriptscriptstyle{(i)}}^{\nu\alpha}$
in the momentum space and to linearize the expressions for $\Pi_{\scriptscriptstyle{(i)}}^{\mu\nu}$ that we found in the previous sections. 
At first order, we have
\begin{equation}
\delta \Pi_{\scriptscriptstyle{(1)}}^{\mu\nu}=-\zeta\Delta^{\mu\nu}\delta\Theta - \eta\,\delta \sigma^{\mu\nu}, \label{delta-pi1}
\end{equation} 
where the expressions for $\delta \Theta$ and $\delta \sigma^{\mu\nu}$ are given respectively by 
\begin{equation}
\delta\Theta = i k_{_\perp}^\mu\delta u_\mu,\qquad
\delta\sigma^{\mu\nu} = i (k_{_\perp}^\mu \delta u^\nu+k_{_\perp}^\nu \delta u^\mu)
-i\frac{2}{d-1}\Delta^{\mu\nu}k_{_\perp}^\alpha\delta u_\alpha.
\label{pi1-aux}
\end{equation}
Inserting expressions \eqref{delta-pi1} and \eqref{pi1-aux} into Eq. (\ref{dispersion}) and eliminating $\delta \ln s$ by using Eq.~\eqref{longitf}, we find
 \begin{equation}
 \left[-i\eta\frac{k_\perp^2}{\varepsilon +p } + k_\parallel\right]\delta u^\mu
 +  \left\{\frac{1}{\varepsilon +p }\left[\frac{\eta(d-3)+\zeta(d-1)}{d-1}\right]-\frac{c_s^2}{k_{_\parallel}}\right\} 
\left(k_{_\perp}^\alpha\delta u_\alpha\right)ik_\perp^\mu=0, \label{firstorder}
\end{equation}
where the definitions \eqref{kprojections} have also been employed. This is a vector equation involving the 
perturbations  in the velocity field as well as in the wavevector, and we cannot assume that these vectors
are linearly independent. However,  we can take projections into independent directions and, as we are describing
wave phenomena, it is natural to use the wavevector to define such directions.
In fact, we can classify the perturbations as longitudinal (sound)
or transversal (shear) as they propagate in the direction of $k_{\perp}^\mu$
or orthogonal to it. 
So the longitudinal sector is selected by applying on equation \eqref{firstorder}
the projector
\begin{equation}
 \Sigma_\parallel^{\mu\nu}=\frac{k_{_\perp}^\mu k_{_\perp}^\nu}{k_{_\perp}^2}.
\end{equation}
Note that $\Sigma_\parallel^{\mu\nu} k_{\perp \nu} = k_\perp^\mu,$ while
$\Sigma_\parallel^{\mu\nu} \delta u_\nu = k_\perp^\mu (k_{_\perp}^\alpha\delta u_\alpha)/k_\perp^2.$
Hence the dispersion relation for the sound waves involves terms multiplying both $\delta u^\mu$ and
$k_\perp^\mu$, and at first order we have
\begin{equation}
 k_{_\parallel}^2 + i\left[\frac{2\eta(d-2)+\zeta(d-1)}{(\varepsilon +p)(d-1) }\right]k_\parallel k_\perp^2 - c_s^2k_{_\perp}^2 = 0.
\end{equation}
On the other hand, to select the shear-mode sector we use the transverse projector
\begin{equation}
 \Sigma_\perp^{\mu\nu}= \Delta^{\mu\nu} - \frac{k_{_\perp}^\mu k_{_\perp}^\nu}{k_{_\perp}^2},
\end{equation}
and, this time, we have that $\Sigma_{\perp \nu}^{\mu}k_{\perp}^\nu = 0$. So the terms multiplying
$k_\perp^\mu$ do not appear in the dispersion relation and, at first order, we get for the shear mode
that
\begin{equation}
 k_{_\parallel} + i\eta\frac{k_{_\perp}^2}{\varepsilon +p } = 0.
\end{equation} 
 
The equations obtained so far, in this section, are all first order in the gradient expansion, and
their solutions provide consistent approximations for $\omega(k)$ up o the order of $k^2$.
However, our goal here is to study third-order hydrodynamics, 
so now we consider the second- and third-order corrections and then
solve the equations to get $\omega(k)$ up to order of $k^4$. 

After linearization, the second-order corrections get simplified and 
the only terms that survive are the gradients of $\Theta$ and $\sigma^{\mu\nu}$,
yielding 
\begin{equation}
 \delta \Pi_{\scriptscriptstyle{(2)}}^{\mu\nu} = \xi_5\,\Delta^{\mu\nu}u^\alpha\partial_\alpha \delta\Theta + \lambda_6 \,u^\alpha\partial_\alpha\delta\sigma^{\mu\nu}.
\end{equation}
It is seen that the linearization process reduces a list of six scalars and eight tensors to one scalar and one tensor.
The effect of additional derivatives is easily taken into account. In the Fourier space, it is a task
of multiplying the respective terms by powers of the spatial wavevector. 
Then, the second-order equation for the shear mode results in
\begin{equation}
 k_{_\parallel} + i\eta\frac{k_{_\perp}^2}{\varepsilon +p } +\lambda_6\frac{k_{_\perp}^2k_{_\parallel}}{\varepsilon +p} = 0,
\end{equation}
while for the sound-wave mode it gives
\begin{equation}
 k_{_\parallel}^2 - c_s^2k_{_\perp}^2 +i\frac{1}{ \varepsilon +p }\left[\zeta +2\eta\left(\frac{d-2}{d-1}\right) \right]k_{_\parallel} k_{_\perp}^2
-\frac{1}{ \varepsilon +p }\left[\xi_5 +2\lambda_6\left(\frac{d-2}{d-1}\right) \right]k_{_\parallel}^2 k_{_\perp}^2= 0.
\end{equation}

Finally, we proceed in the same way to include third-order corrections in the linearized
equations of motion. 
In comparison to the second-order terms, this time we have one more possible structure.
The Laplacians $\partial^2\Theta$ and $\partial^2\sigma $ and combined derivatives of the expansion,
$\partial \partial \Theta$, appear in the linearized energy-momentum tensor, so that the third-order correction reads
\begin{equation}
\delta \Pi_{\scriptscriptstyle{(3)}}^{\mu\nu}= \left(\chi_1 - \frac{2\gamma_{14}}{d-1}\right)\Delta^{\mu\nu}\partial_{_\perp}^2 \delta\Theta
 +\gamma_{13}\partial_{_\perp}^2 \delta\sigma^{\mu\nu}+2\gamma_{14}\partial_{_\perp}^{\langle\mu} \partial_{_\perp}^{\nu\rangle}\delta\Theta, \label{deltat3}
\end{equation}
where $\partial_{_\perp}^\mu = \Delta^{\mu\nu}\partial_\nu.$ In the Fourier space, the derivative
$\partial_{\perp}^\mu$ is replaced by $ik_{\perp}^\mu$, and then the third-order
dispersion relation for the shear mode becomes
\begin{equation}
 k_{_\parallel} -i\eta\frac{k_{_\perp}^2}{\epsilon +p } +\lambda_6\frac{k_{_\perp}^2k_{_\parallel}}{\epsilon +p}-
 i\gamma_{13}\frac{k_{_\perp}^4}{\varepsilon +p} = 0, \label{3shear}
\end{equation}
while for the sound-wave mode it is
\begin{align}
 k_{_\parallel}^2 - c_s^2k_{_\perp}^2 +&i\frac{1}{\varepsilon+p}\left[\zeta +2\eta\left(\frac{d-2}{d-1}\right) \right]k_{_\parallel} k_{_\perp}^2 \nonumber \\
-\frac{1}{ \varepsilon +p }\left[\xi_5 +2\lambda_6\left(\frac{d-2}{d-1}\right) \right]k_{_\parallel}^2 k_{_\perp}^2 
&-\frac{i}{ \varepsilon +p }\left[\chi_1 +2(\gamma_{13}+\gamma_{14})\left(\frac{d-2}{d-1}\right) \right]k_{_\parallel} k_{_\perp}^4= 0. \label{3sound}
\end{align}

It is notorious that the holographic calculation of $\eta/s$ for $\mathcal{N}=4$ SYM
is one of the strongest results from the AdS/CFT correspondence and represents
a particular case where the study of linear perturbations in
holography provides a full description of the system,
since at first order a conformal fluid has only one transport coefficient.
Unfortunately, this is not the case for higher-order hydrodynamics, where the linearized equations
involve a very reduced number of transport coefficients obtained from all the corresponding higher-order terms.
The present analysis allows us to evaluate only the transport coefficients associated with structures that survive to linearization.

\subsection{Dispersion relations in the comoving (static) frame}
\label{Sec:DispersionFluids}

Due to relativistic effects the dispersion relation must be defined in a fixed reference frame,
where the notion of frequency is meaningful. We look first to the static frame where the observer sees no flow,
measuring the fluid velocity $u^{\mu}_{\mbox{\scriptsize{static}}}=(1,\, \vec{0})$.  
In such a situation, $\Delta^\mu_{~\nu}$ projects a vector on the hypersurfaces of constant time,
since  $\Delta^\mu_{~\nu}=\delta^\mu_{~\nu} + \delta^\mu_{~0}\,\eta_{\nu 0} = {\rm diag}\left(0,\,1,\,1,\,1\right).$
As a consequence, the parallel and perpendicular components of $k^\mu$ are precisely the
frequency and the space part of the wavevector,
\begin{equation}
  k_{_\parallel} = u_\mu k^\mu = -\omega,\qquad k_{_\perp}^\mu=  (0,\, \vec{k}), \label{static}
\end{equation}
and hence $k_\perp^2=k_\perp^\mu k_{\perp \mu}=\vec{k}\cdot\vec{k}\equiv k^2$.

We build the dispersion relations by starting with the shear-mode case.
Inserting the expressions from Eq.~\eqref{static} into Eq.\eqref{3shear},
we find an algebraic equation for $k$ and $\omega$ that, up to fourth-order corrections,
is given by  
\begin{equation}
 -\omega -i\frac{\eta}{\varepsilon +p}k^2 - \frac{\lambda_6}{\varepsilon +p}
 \omega k^2 +i\frac{\gamma_{13}}{\varepsilon +p}k^4=0,
\end{equation}
whose solution for $\omega(k)$ furnishes the dispersion relation, 
\begin{equation}\label{Eq:DespersionShearFluid}
   \omega_{\mbox{\scriptsize{shear}}}(k)= -i\frac{ \eta}{s\mathcal{T} }  k^2  -i
   \left(\frac{\eta \lambda_6}{s^2\mathcal{T}^2}- \frac{\gamma _{13}}{s\mathcal{T}}  \right) k^{4}, 
\end{equation}
where we made use of the equilibrium thermodynamics to write $p+\epsilon= s\mathcal{T}$.

To deal with the sound-wave mode it is convenient to define reduced coefficients by
the combination of transport coefficients that appear in the expressions for
the dispersion relations as 
\begin{equation}
 \beta_1= \zeta +2\eta\frac{d-2}{d-1}, \qquad \beta_2 = \xi_5 +2\lambda_6\frac{d-2}{d-1},\qquad \beta_3  = \chi_1 +2(\gamma_{13}+\gamma_{14})\frac{d-2}{d-1}.
\end{equation} 
Using these definitions and substituting the expressions from Eq.~\eqref{static} into Eq.\eqref{3sound},
we get the following algebraic equation  
\begin{equation}
 \omega^2-c^2k^2-i\frac{k^2\omega}{s\mathcal{T}}\beta_1-\frac{k^2\omega^2}{s\mathcal{T}}\beta_2 +i\frac{k^4\omega}{s\mathcal{T}}\beta_3=0.
\end{equation} 
This is a polynomial of second order in $\omega$, which generates two solutions, $\omega_+$ and $\omega_-$,
corresponding to waves traveling in opposite directions: $\omega_+ $ ($\omega_- $) corresponds to modes
propagating forwards (backwards) with respect to the wavevector direction.
The power series solution for the $\omega_+$ mode up to $k^4$ gives 
\begin{equation}\label{Eq:DispersionSoundFluid}
\omega_{+}(k)=ck-\frac{i \beta_1}{2s\mathcal{T}}k^2
-\left(\frac{\beta_1^2}{8\,c\,s^2\mathcal{T}^2}-\frac{c\, \beta_{2}}{2s\mathcal{T}}\right)k^3
-i\left(\frac{\beta_3}{2s\mathcal{T}}-\frac{\beta_1\,\beta_2}{2s^2\mathcal{T}^2}\right)k^4.
\end{equation} 
The negative solution $\omega_-$  is related to the positive one by $\omega_-(k)=\omega_+(-k)$.

The resulting expressions for the dispersion relations involve the transport coefficients
directly related with the shear tensor in Eq. (\ref{Eq:DespersionShearFluid}) and combinations of transport coefficients either from shear and expansion appearing in Eq. (\ref{Eq:DispersionSoundFluid}) in the form of the reduced coefficients 
$\beta_1$, $\beta_2$ and $\beta_3.$ The dispersion relations in the presence of conformal symmetry are obtained
from the general, non-conformal, case by imposing constrains on the transport coefficients.
For the choices we performed here we obtain dispersion relations for the conformal fluid by
fixing  $\zeta=\xi_5=0,~\lambda_6=\upsilon_4$ at first and second order, while at third order we have that $\gamma_{14}=\vartheta_6,~\gamma_{13}=\vartheta_7,~\chi_1=0$ .
The transport coefficient $\vartheta_{11}$ is absent in the dispersion relation of a conformal fluid because the linear fluctuation of the tensor $u_\eta\mathcal{D}^{\langle\mu}\mathcal{F}^{\nu\rangle\eta}$ in a static flow vanishes. The identification gives the expressions for the reduced coefficients present in the sound-wave mode as functions of the transport coefficients of conformal hydrodynamics 
\begin{equation}
\beta_1=2\eta\left(\frac{d-2}{d-1}\right),\qquad \beta_2=2\upsilon_4\left(\frac{d-2}{d-1}\right),\qquad \beta_3=2(\vartheta_6+\vartheta_7)\left(\frac{d-2}{d-1}\right).    
\end{equation}

\subsection{Dispersion relations in a moving frame} 

In a moving frame the velocity flow may be decomposed as $u^\mu = \gamma (1,\,\vec{v})$, with $\gamma=1/\sqrt{1-v^2}$, and the projections $k_\parallel$ and $k_\perp ^\mu$ defined in Eq.~\eqref{kprojections} are such that
\begin{equation}\label{kmoving}
 k_\parallel = \gamma(\vec{k}\cdot\vec{v} -\omega),\qquad k_\perp^2 = -\omega^2 + \vec{k}^{\,2} + \gamma^2(\vec{k}\cdot\vec{v} -\omega)^2.
\end{equation}
Here it is convenient to project the space part of the wavevector in the parallel and perpendicular
directions with respect to the ordinary velocity $\vec v$
by defining $\vec{m} = (\vec{k}\cdot\hat{v})\hat{v}$ and $\vec{q} = \vec{k} - \vec{m}$.
From these definitions, it follows that $\vec{k}\cdot\vec{v}=mv$ and $\vec{k}^{\,2} = q^2+m^2$, and 
Eq.~\eqref{kmoving} gives  
\begin{equation}
k_\parallel = \gamma(-\omega +mv), \qquad k_\perp^2 = q^2+\gamma^2(m-\omega v)^2.
\end{equation}
Inserting the relations from the last equation into Eqs.~(\ref{3shear}) and (\ref{3sound}), we find, respectively,
\begin{align} 
i\frac{\gamma_{13}}{s\mathcal{T}} \left(\frac{(m-v \omega )^2}{1-v^2}+q^2\right)^2-&i\frac{\eta}{s\mathcal{T}}\left(\frac{(m-v \omega )^2}{1-v^2}+q^2\right) \nonumber \\
+&\frac{\lambda_6}{s\mathcal{T}} \frac{m v-\omega}{\sqrt{1-v^2} } \left(\frac{(m-v \omega )^2}{1-v^2}+q^2\right)+\frac{m v-\omega }{\sqrt{1-v^2}}=0,
\end{align}
for the shear mode, and
\begin{align}
 \frac{(m v-\omega )^2}{1-v^2}-c^2\sqrt{\frac{(m-v \omega )^2}{1-v^2}+q^2}+& i\frac{ \beta_1}{{s\mathcal{T}}} 
 \frac{m v-\omega }{\sqrt{1-v^2}} \sqrt{\frac{(m-v \omega )^2}{1-v^2}+q^2}\nonumber \\
 -\frac{ \beta_2}{{s\mathcal{T}}}  \frac{(m v-\omega)^2 }{1-v^2}\sqrt{\frac{(m-v \omega )^2}{1-v^2}+q^2}&-i\frac{ \beta_3}{{s\mathcal{T}}}  \frac{m v-\omega }{\sqrt{1-v^2}}\left(\frac{(m-v \omega )^2}{1-v^2} +q^2\right)=0,
\end{align}
for the sound-wave mode.
The dispersion relations in the moving frame $\omega = \omega(m,q)$ are given implicitly
by the last two equations. Explicit  solutions for $\omega$ in power series of $m$ and $q$
result in cumbersome expressions for the general case and are not presented here.

\section{The $\mathcal{N}=8$ SUSY Yang-Mills plasma} 
\label{Sec:Application}

\subsection{General remarks}

In this section we apply the general results presented above to the study of 
a system that can be holographically described by the $\mbox{AdS}_{4}/\mbox{CFT}_3$
correspondence \cite{Maldacena:1997re}, namely, the $\mathcal{N}=8$ supersymmetric (SUSY) 
Yang-Mills plasma\footnote{It is worth pointing out that the world-volume theory of $N$
M2-branes \cite{Townsend:1995af} is equivalent to a $\mathcal{N}=8$ 
super Yang-Mills theory in $(2+1)$-dimensions, and the plane-symmetric black holes 
can be obtained in the near horizon approximation of the supergravity action 
(see Ref. \cite{Herzog:2002fn} for details). This theory may also be considered as a 
particular case of the ABJM theory \cite{Aharony:2008ug}.}.
Specifically, we deal with the problem from the gravitational point of view,
in order to obtain the transport coefficients of the plasma up to third order
in the gradient expansion. As it is known from the holographic dictionary,
the metric of an asymptotically AdS spacetime is dual to the energy-momentum tensor
of the boundary conformal field theory. This allows us to relate perturbations of a
classical black-hole background to the energy-momentum fluctuations in the dual thermal field theory.
We focus in the long-distance and low-energy behavior of these perturbations, i.e.,
the regime where the constituents of the system present the collective behavior typical
of a continuous medium \cite{Landau:1987,Kovtun:2012rj}. 
In the context of the gauge/gravity duality, this is the hydrodynamic limit of the 
gravitational perturbations \cite{Policastro:2002se,Policastro:2002tn}.

The gravitational AdS background is a plane-symmetric static black hole, which 
is a solution of the Einstein equations with negative cosmological 
constant obtained in Ref.~\cite{Lemos:1994xp}. To deal with this 
problem we need to extend the previous results presented 
in Refs.~\cite{Miranda:2005qx,Miranda:2008vb,Morgan:2009pn},
since they were obtained using a first-order expansion of the energy-momentum 
tensor. In order to obtain the dispersion relations, we are going to use the 
techniques implemented in Refs.~\cite{Ge:2010yc,Brattan:2010pq}.
The resulting transport coefficients are obtained by comparing the dispersion 
relations obtained in both sides of the AdS/CFT duality.

\subsection{The background spacetime and the perturbation equations} 
The gravitational background considered here is the spacetime of a
plane-symmetric static AdS$_4$ black hole whose metric may be written
in the form \cite{Lemos:1994fn}
\begin{equation}\label{background1} 
ds^2= {\alpha^2r^2} \left[-f(r)dt^2+dx^2+dy^2\right]
+\frac{dr^2}{\alpha^2r^2 f(r)}, 
\end{equation}
where $\alpha$ is a parameter related to the AdS radius $\ell$ by $\alpha = 1/\ell=\sqrt{-3/\Lambda}$, with $\Lambda$ being
the negative cosmological constant, and $f(r)=1-r_h^3/r^3$ the horizon function (also named the redshift factor, or the blackening 
factor). 
The coordinate ranges are $-\infty< t< +\infty$, $0\leq r < \infty$, 
$-\infty< x < +\infty$, and $-\infty < y < +\infty$. This metric describes
a black hole spacetime whose event horizon is located at $r=r_h$, where $f(r)=0$,
and the AdS boundary lies at $r\to \infty$, where $f(r)\to 1$. Since the topology
of the surfaces $t,\, r$ = constant is planar, the metric \eqref{background1}
is usually referred as describing black branes.  

The Hawking temperature of static black holes (and black branes) is 
obtained by evaluating the following expression at the horizon
\noindent
\begin{equation}
\mathcal{T}=\frac{1}{4\pi}
\Bigg{|}\frac{d}{dr}\left(\alpha^2r^2\,f(r)\right)\Bigg{|},
\end{equation}
so that we obtain 
\noindent
\begin{equation}\label{TemperatureBS}
\mathcal{T}=\frac{3\alpha^{2}r_h}{4\pi}.
\end{equation}
\noindent
 
\noindent
From the dual field-theory perspective, $\mathcal{T}$ is the local temperature of the 
conformal plasma \cite{Bhattacharyya:2007vs,Caldarelli:2008ze}.
For the forthcoming analysis it is convenient to 
introduce the variable $u=r_h/r$. In such a coordinate the AdS boundary 
lies at $u=0$ and the horizon at $u=1$. To completely characterize the AdS
black hole as a thermodynamic system we have to consider the equation of 
state $\epsilon=-p+s\,\mathcal{T}$, where $\epsilon$, 
$p$, and $s$ are the energy, pressure and entropy densities, 
respectively. The entropy is determined through the Bekenstein-Hawking formula
\noindent
\begin{equation}\label{Eq:Entropy}
S=\frac{\mathcal{A}}{4\,G},
\end{equation}
\noindent
where $\mathcal{A}$ is the horizon area and $G$ is the gravitational 
coupling. There is a relationship between the gravitational coupling $G$
and the number of colours $N$ \cite{Klebanov:1996un} that appears in the context of the
AdS/CFT correspondence. This relation read as $G=3\,\ell^2/(2\,N)^{3/2}$.
In turn, the horizon area is given by $\mathcal{A}=r_h^{2}\,V_2/\ell^2$, where $V_2$ represents the 
``volume'' along the transverse coordinates. Hence, after replacing
it in \eqref{Eq:Entropy} and using \eqref{TemperatureBS} we get
\noindent
\begin{equation}\label{Eq:Entropy2}
s=\frac{S}{V_2}=\frac{2^{7/2}}{3^3}\pi^2\,N^{3/2}\,\mathcal{T}^{2}.
\end{equation}
\noindent
This result is in agreement with the one obtained in Ref. \cite{Klebanov:1996un}.

It is worth mentioning that gravitational perturbations of static plane-symmetric black holes
were extensively investigated in the literature; see, for instance, 
Refs.~\cite{Miranda:2005qx,Miranda:2008vb,Morgan:2009pn}. 
The perturbations of the rotating and electrically charged counterparts
to the black branes were also studied in details \cite{Miranda:2014vaa, Mamani:2018qzl}.
In this part of the paper we are going to use the limit of zero charge and zero angular momentum of the results
presented in Ref.~\cite{Miranda:2014vaa}.  
The motivation for using these results here 
is to explore the quantum-mechanical SUSY relationship between the vector
and the scalar sectors of the black-brane gravitational perturbations.
We give additional details about this supersymmetry in Sec.~\ref{Ref:SubSecScalarAnalytic}.

Let us start by writing the differential equations governing the vector and the scalar sectors 
of the perturbations on the background 
\eqref{background1}.\footnote{Let us stress that the vector sector is also known as the shear mode,
while the scalar sector as the sound mode. This nomenclature is justified because, in the hydrodynamic limit,
their dispersion relations are the same as the shear diffusion and sound propagation modes, 
respectively.} As remarked in Ref.~\cite{Miranda:2014vaa}, in the zero charge limit, 
the gravito-electromagnetic perturbations decouple and may be split in electromagnetic 
and gravitational perturbations. The interest here is to investigate the gravitational perturbations, which 
are represented by the variables $Z_{\scriptscriptstyle{2}}^{(-)}$ and 
$Z_{\scriptscriptstyle{2}}^{(+)}$ for the vector and scalar sectors, respectively. These variables
are known in the literature as the Regge-Wheeler-Zerilli (RWZ) master variables.
In order to compare with previous results 
\cite{Miranda:2005qx,Miranda:2008vb,Morgan:2009pn,Mamani:2018qzl},
we change the notation for the master variables, 
$Z_{\scriptscriptstyle{2}}^{(-)} \to \Phi_{\scriptscriptstyle{V}}$ and 
$Z_{\scriptscriptstyle{2}}^{(+)} \to \Phi_{\scriptscriptstyle{S}}$.
Hence, the  Schr\"odinger-like form of the differential equations 
become (cf. Eq.~(71) in \cite{Miranda:2014vaa})
\begin{equation}\label{eqRWZ}
\left[f\partial_{u}\left(f\partial_{u}\right)+ \mathfrak{w}^2-
V_{\scriptscriptstyle{V,S}}
\right]\Phi_{\scriptscriptstyle{V,S}}=0,
\end{equation}
where 
$V_{\scriptscriptstyle{V}}$ and $V_{\scriptscriptstyle{S}}$ 
refer, respectively, to the vector and scalar potentials, 
defined as 
\begin{align}
V_{\scriptscriptstyle{V}}(u)=&f\left(\mathfrak{q}\,^2 -3u\right),
\label{potential_T}\\ 
V_{\scriptscriptstyle{S}}(u)=&\dfrac{f}{\mathfrak{q}\,^2 +3u}
\left[\mathfrak{q}^4 + \dfrac{9\,\big(2+\mathfrak{q}\,^2 
u^2+u^3\big)}{\mathfrak{q}\,^2+3\, u}\right],
\label{potential_L}
\end{align}
where $\mathfrak{w}$ and $\mathfrak{q}$ are the frequency and the magnitude of the
wavevector, both normalized by the Hawking temperature $\mathcal{T}$,
\begin{equation} \label{eq:norm}
\mathfrak{w}=\frac{3\,\omega}{4\pi \mathcal{T}},\qquad 
\mathfrak{q}=\frac{3\,k}{4\pi \mathcal{T}}. 
\end{equation}
In the following we investigate some of the solutions of Eq.~\eqref{eqRWZ}.

\subsection{Analytical solutions for the vector (or shear) sector} \label{Sec:PerturbVectorSector}

The aim of this part of the work is to generalize the results obtained for
the dispersion relations of vector perturbation modes in four-dimensional spacetimes
(see Ref.~\cite{Miranda:2008vb}), and also in arbitrary $d$-dimensional 
black branes, Ref.~\cite{Morgan:2009pn}. Previous results 
in the literature were obtained by considering first- and second-order hydrodynamics
(see e.g. Refs.~\cite{Natsuume:2007ty,Baier:2007ix}); here we go up to the third order.

Let us first investigate the behavior of the function 
$\Phi_{\scriptscriptstyle{V}}(u)$ close to the horizon, i.e., for $u$ in the vicinity of $u=1$. 
In such a region, 
the wave function behaves like $\Phi{\scriptscriptstyle{V}}\propto f^{\,\beta}$, 
where $\beta$ is a constant parameter that may be obtained by replacing 
this expression into Eq.~\eqref{eqRWZ}. In doing so, the leading-order equation  
has two solutions for $\beta$, namely, $\beta=\pm i\mathfrak{w}/3$, 
where $- i\mathfrak{w}/3$ corresponds to an ingoing wave 
falling into the black-hole horizon, while the other solution
corresponds to an outgoing wave. In the forthcoming analysis we use the 
ingoing solution at the horizon, because 
classically nothing comes out of the black-hole interior.

In the next stage, we are going to solve Eq.~\eqref{eqRWZ} 
perturbatively in the parameters $(\mathfrak{w}, \mathfrak{q}^2)$.
Inspired by the procedure implemented in Refs.~\cite{Ge:2010yc,Brattan:2010pq},
we assume that the wave function $\Phi_{\scriptscriptstyle{V}}$ may be expressed as follows
\noindent
\begin{equation}\label{EqPhiVec}
\Phi_{\scriptscriptstyle{V}}(u)=
f^{-i\mathfrak{w}/3}F(u,\mathfrak{w},\mathfrak{m},\mathfrak{q})G(u).
\end{equation}
\noindent
By plugging this into Eq.~\eqref{eqRWZ} it drives to two second-order differential equations, one for $G(u)$ and another for $F(u,\mathfrak{w},\mathfrak{m},\mathfrak{q})$. It is possible to split the equations so that $G$ is an arbitrary function of the radial holographic coordinate only, while $F$ depends on $u$ and also on the frequency  $\mathfrak{w}$ and wavenumbers $(\mathfrak{m},\mathfrak{q})$. Additionally, it is required that $ F(u,\mathfrak{m},\mathfrak{q})$ reduces to a constant 
in the limit $(\mathfrak{m},\mathfrak{q})\to(0,0)$. For more details on this procedure see Appendix~\ref{Sec:AppendixA}.
Hence, the transformation \eqref{EqPhiVec} allows us to simplify drastically 
the analysis of solving Eq.~\eqref{eqRWZ}. The next step is to expand $F(u,\mathfrak{m},\mathfrak{q})$ as a double
series in the parameters $\mathfrak{w}$ and $\mathfrak{q}$. To capture results up to third-order in the hydrodynamic approximation, 
we expand $F(u)$ up to fourth-order
\cite{Policastro:2002se,Policastro:2002tn},
\noindent
\begin{equation}\label{EqExpansionF}
F(u,\mathfrak{w},\mathfrak{m},\mathfrak{q})=F_0+\mathfrak{w} F_1+\mathfrak{q}^2F_2+
\mathfrak{w}^2F_3+\mathfrak{w}\,\mathfrak{q}^2F_4+
\mathfrak{q}^4F_5+\cdots,
\end{equation}
\noindent
where $F_0$ is a constant, while  $F_1, F_2,\cdots$ depend on the 
holographic coordinate and are determined by solving order by order
the resulting differential equation for $F(u,\mathfrak{m},\mathfrak{q}) $,
see Eq.~\eqref{eq:forFu}.

Once the wave function \eqref{EqPhiVec} is obtained, it is expanded close to the boundary, $u=0$, 
in a power series of $u$ as 
\noindent
\begin{equation}\label{EqVectorBound}
\Phi_{\scriptscriptstyle{V}}(u)=
\widetilde{\Phi}_{\scriptscriptstyle{V}}
\left(1+\Pi_{\scriptscriptstyle{V}}\,u+\cdots\right),
\end{equation}
\noindent
where $\widetilde{\Phi}_{\scriptscriptstyle{V}}$ and $\Pi_{\scriptscriptstyle{V}}$ are functions 
of the parameters $(\mathfrak{w}$ and $\mathfrak{q})$,
and the ellipsis represent higher-order contributions. The dispersion relation 
of this sector is identified from the pole of the scalar
function $\Pi_{\scriptscriptstyle{V}}$ \cite{Ge:2010yc,Brattan:2010pq}. The 
explicit form of this function is presented in Appendix~\ref{Sec:AppendixA}.
Here we write the resulting dispersion relation of the vector sector in the hydrodynamic limit, 
\noindent
\begin{equation}\label{EqDispVector}
\omega=-\frac{i}{4\pi\mathcal{T}}{k}^2-
i\frac{9+\sqrt{3}\,\pi-9\,\ln{3}}{384\,\pi^3\mathcal{T}^3}\,k^4
+\mathcal{O}(k^6),
\end{equation}
\noindent
where $\mathcal{O}(k^6)$ indicates higher-order contributions.  
From the gravitational point of view, Eq. \eqref{EqDispVector} represents 
the dispersion relation of the 
low-lying quasinormal perturbation mode of the background \eqref{background1}. 
As we are considering the hydrodynamic limit, the present results may be compared to the
corresponding results obtained from the third-order relativistic hydrodynamics, 
Eq.~\eqref{Eq:DespersionShearFluid}. 
The first term is the leading term of the vector sector and may be compared, in the dual description, with the
result obtained using the first-order hydrodynamics (see Section~\ref{Sec:FirstSecondHydro}).
Moreover, the second-order hydrodynamic contributions vanish, because 
there is no contribution from terms proportional to $k^3$ in Eq.~\eqref{EqDispVector}. However, the subleading 
contribution to Eq.~\eqref{EqDispVector} is due to the fourth-order in the wavenumber term, which may be compared 
with the corresponding results obtained in the third-order hydrodynamics 
(see Section~\ref{Sec:ThirdHydro}). 
As it is clear from the dispersion relation~\eqref{EqDispVector}, the 
frequency is a purely imaginary number representing damping waves. The
corresponding damping time is related to the imaginary part 
$\tau=1/\omega_I$. It is worth mentioning that the same result 
was obtained in Ref.~\cite{Natsuume:2007ty}\footnote{See the SAdS$_4$ case
in Table III to compare with. Notice that the normalization factors are different from the present work. 
The relations between parameters used here and in Ref.~\cite{Natsuume:2007ty} 
are: $\mathfrak{w} =(3/2)\mathfrak{w}^{\scriptscriptstyle{\text{Natsuume}}}$ 
and $\mathfrak{q} =(3/2)\mathfrak{q}^{\scriptscriptstyle{\text{Natsuume}}}$.}.

\subsection{Analytical solutions for the scalar (or sound) sector} 
\label{Ref:SubSecScalarAnalytic}

It is known that obtaining the dispersion relation of the scalar-sector
gravitational perturbations is, in general, an involved and lengthy process. 
To work around this, we use here the quantum-mechanical supersymmetry (SUSY),
also called electric/magnetic duality in the linearized gravity,
that connects the vector and scalar 
sectors of the gravitational perturbations in four-dimensional 
spacetimes. In the case of charged rotating black strings in asymptotically AdS spacetimes,
this supersymmetric relation was investigated  in Ref.~\cite{Miranda:2014vaa} (SUSY 2015). 
It should be clear that the supersymmetric 
relationship arises when the perturbations are represented by the RWZ variables. 

\begin{figure}[htb]
\begin{center} 
\includegraphics[width=14.0cm]{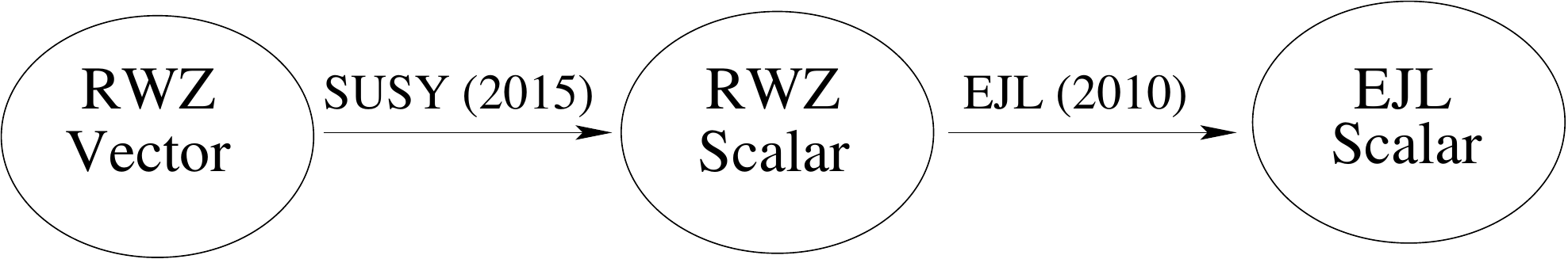}
\end{center}
\caption{Schematic representation of the connection between 
the Regge-Wheeler-Zerilli (RWZ) and Edalati-Jottar-Leigh (EJL) variables.}
\label{Fig:SUSY}
 
\end{figure}

Now it is important to point out the difference between the RWZ and the Kovtun-Starinets (KS) variables, that are frequently used in holography. The KS variables prevail in holography because the imposition of Dirichlet boundary condition on these variables, at the AdS boundary $u=0$, leads to the poles of the retarded Green functions in the corresponding dual field theory. This means that the poles of the retarded 
Green functions, in the momentum space, are equal to the frequencies of the black-brane
quasinormal modes in the gravitational side of the correspondence. 
Hence, the natural way would be to use the relation 
between RWZ and KS variables, as it was obtained in Ref.~\cite{Morgan:2009pn}. 
However, such a relationship is quite involved and we do not employ it here.
A more interesting alternative is adopt another
set of master fields, that were introduced originally in Ref.~\cite{Edalati:2010pn}.
We call these new master fields as the Edalati-Jottar-Leigh (EJL) variables and establish a
relation between them and the RWZ variables.
It is important to mention that KS and EJL variables are equivalent in the sense that both
furnish the poles of the retarded correlation functions in the dual field theory by imposing
Dirichlet conditions on these variables at the AdS boundary.
The reason to employ the EJL set of variables is that the
relation between RWZ and EJL variables is simpler than the relation
between RWZ and KS variables. 
In Fig.~\ref{Fig:SUSY}, we display a schematic representation of 
the road we follow to obtain the dispersion relations of the 
scalar sector by using the EJL variable for the scalar perturbations
and the quantum-mechanical SUSY.
\noindent
\noindent
 
Regarding the difference between the RWZ and EJL variables, 
we point out that the low-lying quasinormal modes, so-called hydrodynamic QNMs,
arise by imposing Robin-type boundary condition on the RWZ variable for the
scalar sector, while for EJL master variable it is necessary just Dirichlet condition at the boundary $u=0$.

To implement the above-mentioned strategy we start by writing the 
supersymmetric relation between the RWZ variables of the vector and scalar sectors 
of the gravitational perturbations.
 In the limit of zero electric charge, the case of interest here, Eqs.~(105) and (106) from Ref.~\cite{Miranda:2014vaa} can be written as\footnote{We point 
out some typos in Ref.~\cite{Miranda:2014vaa}. The corrected version of Eq.~(30) is obtained by replacing $k\to \ell k$.
The correct expression of the operator
$\Lambda^2$ presented in Eq.~(63) is $\Lambda^2=\partial^{2}_{r_{*}} +\varpi^2$.} 
\begin{equation}
\begin{aligned}\label{EqRWZVecToSca}
\left(\ell^4k^4+6i\omega\,\frac{r_h^3}{\ell^2}\right)
\Phi_{\scriptscriptstyle{V}}=&\left(\ell^4k^4+
\frac{18\,r_h^6\,f}{\ell^4\left(k^2\ell^4+3u\,r_h^2\right)}\right)
\Phi_{\scriptscriptstyle{S}} +6\frac{r_h^4\,f}{\ell^4}\partial_{u}\Phi_{\scriptscriptstyle{S}},
\end{aligned}
\end{equation}
\begin{equation}
\begin{aligned}\label{EqRWZScaToVec}
\left(\ell^4k^4-6i\omega\,\frac{r_h^3}{\ell^2}\right)
\Phi_{\scriptscriptstyle{S}}=&\left(\ell^4k^4+
\frac{18\,r_h^6\,f}{\ell^4\left(k^2\ell^4+3u\,r_h^2\right)}\right)
\Phi_{\scriptscriptstyle{V}} -6\frac{r_h^4\,f}{\ell^4}\partial_{u}\Phi_{\scriptscriptstyle{V}},
\end{aligned}
\end{equation}

The next stage is to write explicitly the
relation between $\Phi_{\scriptscriptstyle{S}}$ and EJL variable. 
For our benefit this relation was obtained in Ref.~\cite{Edalati:2010pn} and is given by 
\noindent
\begin{equation}\label{EqEJLScaToSca}
\psi_{\scriptscriptstyle{S}}=\frac{3r_h^2}{k^2\,\ell^4}
\Phi_{\scriptscriptstyle{S}}+
f\,\partial_{u}\Phi_{\scriptscriptstyle{S}}.
\end{equation}
\noindent
Hence, $\psi_{\scriptscriptstyle{S}}$ can be written in terms of the
wave function $\Phi_{\scriptscriptstyle{V}}$. The EJL 
variable $\psi_{\scriptscriptstyle{S}}$ may be completely determined in the hydrodynamic limit,
because $\Phi_{\scriptscriptstyle{V}}$ is given by Eq.~\eqref{EqPhiVec}.
It is possible to find an explicit relation between $\psi_{\scriptscriptstyle{S}}$ and 
$\Phi_{\scriptscriptstyle{V}}$ if we replace Eq.~\eqref{EqRWZScaToVec} in 
Eq.~\eqref{EqEJLScaToSca}. Additionally, we should introduce the 
normalized parameters $\mathfrak{w}$ and $\mathfrak{q}$, which are
defined as $\omega=\mathfrak{w}\,r_h/\ell^2
=\mathfrak{w}\,r_h\alpha^2$ and $k=\mathfrak{q}\,r_h/\ell^2
=\mathfrak{q}\,r_h\alpha^2$.
At the end we obtain 
\noindent
\begin{equation}\label{EqScalarVector}
\begin{split}
\mathfrak{q}^2\left(\mathfrak{q}^4 -6i\mathfrak{w}\right)
\psi_{\scriptscriptstyle{S}}=&3\Big(
\mathfrak{q}^2\left[\mathfrak{q}^2-2\mathfrak{w}^2\right]
-2\,r_h^2\,\mathfrak{f}_2\left[\mathfrak{q}^6 -9\left(1-
\mathfrak{q}^2r_h^2\,\mathfrak{f}_2\right)\right]\Big)
\Phi_{\scriptscriptstyle{V}}\\
&+f\Big(\mathfrak{q}^6 -18\left[1-\mathfrak{q}^2r_h^2\,\mathfrak{f}_2\right]
\Big) \partial_{u}\Phi_{\scriptscriptstyle{V}}.
\end{split}
\end{equation}
\noindent
This is the full relation between the vector sector, described 
by the RWZ variable, and the scalar sector, described by using the EJL variable.  
We then obtain the dispersion relation of the low-lying scalar QNMs 
by imposing Dirichlet condition on the master 
field $\psi_{\scriptscriptstyle{S}}$ at the boundary (for discussions 
see, for instance, Refs.~\cite{Morgan:2009pn,Michalogiorgakis:2006jc,
Dias:2013sdc,Mamani:2018qzl}). The resulting dispersion relation of the scalar sector
may be written as
\noindent
\begin{equation}\label{EqDispScalar}
\begin{split}
\omega=&\frac{1}{\sqrt{2}}k -\frac{i}{8\pi\,\mathcal{T}}k^2
+\frac{\left(15+\sqrt{3}\,\pi-9\ln{3}\right)}{192\,\sqrt{2}\,\pi^2\,\mathcal{T}^2}k^3\\
&+\frac{i\big[144+4\,\sqrt{3}\,\pi+3\,\pi^2 -6\left(6+\sqrt{3}\,\pi\right)\ln{3}+27\ln^2{3}
-12\,\psi^{(1)}\left({2}/{3}\right) \big]}{6144\,\pi^3\,\mathcal{T}^3}k^4\\
&+\mathcal{O}(k^5),
\end{split}
\end{equation}
\noindent
where $\psi^{(1)}\left({2}/{3}\right)$ is the polygamma function
and $\mathcal{O}(k^5)$ represents higher-order contributions.  
Eq.~\eqref{EqDispScalar} extends previous results in the literature. For instance, 
by comparing it with the result of 
Ref.~\cite{Natsuume:2007ty} (see Table IV, the $SAdS_4$ case, in this reference)
we observe that both results are in agreement up to third order in the wavenumber.
This is a remarkable result since we found 
Eq.~\eqref{EqDispScalar} by using as starting point the results obtained 
for the vector sector, and then we apply the SUSY relationship.
Moreover, we point out that the last term on the right-hand-side of Eq.~\eqref{EqDispScalar}
was not previously obtained in the literature. We also notice that the
dispersion relations of the scalar sector have real and 
imaginary parts, representing waves propagating and being damped, 
differently from the wave functions of the vector sector, which 
are purely damped.

\subsection{Transport coefficients}

After obtaining the dispersion relations 
of the gravitational perturbations in the hydrodynamic limit, cf.
Eqs.~\eqref{EqDispVector} and \eqref{EqDispScalar}, we are able to get the transport coefficients by 
comparing such relations to the corresponding dispersion relations found 
in the relativistic hydrodynamic analysis, cf. Eqs.~\eqref{Eq:DespersionShearFluid} 
and \eqref{Eq:DispersionSoundFluid}, respectively.
To determine the transport coefficients we also must consider 
the conformal case with $d=3$, where $\beta_1=\eta$, $\beta_2=\upsilon_4$ and 
$\beta_3=\vartheta_6+\vartheta_7$. So we finally obtain
\noindent
\begin{equation}\label{Eq:DispersionShearGrav}
\begin{aligned}
\eta&=\frac{2^{3/2}\pi}{3^3}N^{3/2}\,\mathcal{T}^{2},\qquad
\upsilon_4=\frac{18+\sqrt{3}\,\pi-9\ln{3}}{162\,\sqrt{2}}N^{3/2}\,\mathcal{T},\qquad
\vartheta_{7}=\frac{1}{72\,\sqrt{2}\,\pi}N^{3/2},\\
\vartheta_6&=\frac{12\psi^{(1)}\left(\frac{2}{3}\right)
-3\pi^2+2\sqrt{3}\,\pi\left(2+3\ln{3}\right)
-9\ln{3}\left(4+3\ln{3}\right)}{5184\,\sqrt{2}\,\pi}N^{3/2}-\vartheta_{7},\\
\end{aligned}
\end{equation}
\noindent
where we have also used Eq.~\eqref{Eq:Entropy2}.
It is worth noticing that the transport coefficients emerging at the second 
order in the gradient expansion depend on the temperature, whereas the 
transport coefficients emerging at third order do not depend 
on $\mathcal{T}$. From these coefficients the relaxation time, 
which is frequently used in the literature, reads
\noindent
\begin{equation}
\tau_{\Pi}=\frac{\upsilon_4}{\eta}=\frac{18+\sqrt{3}\,\pi-9\,\ln{3}}{24\,\pi\,\mathcal{T}}.
\end{equation}

\section{Final remarks}
\label{Sec:Conclusion}

We have faced the program of building the theory of hydrodynamics in order to find out the complete description
of conformal and non-conformal fluids up to third order in gradient expansion. By assuming that hydrodynamics is an effective theory that respects the underlying symmetries of the system we state that the construction of the gradient expansion should include only structures that are completely covariant under these symmetries. For non-conformal fluids we find that the full set of independent
third-order gradients is composed by 19 scalars, 34 vectors and 39 tensors, thus requiring 58 additional transport coefficients. 
This represents a reduction of 10 transport coefficients in comparison with the current literature and is a consequence of the equivalences between structures that differ only by the order of derivatives. 
The presence of conformal symmetry is treated on the same foot as the general covariance and we consider only Weyl covariant gradients in our construction. We find a set of 19 conformal third-order tensors, representing a reduction of one transport coefficient in comparison to the present results in the literature. We also find 6 independent conformal scalars and 11 independent conformal vectors. It is worth mentioning that the second law of thermodynamics is another possible source of constraints for reducing the number of structures (see for instance Ref.~\cite{Glorioso:2016gsa}). This is an interesting subject and we will address it in the future\footnote{We thank the anonymous referee for pointing out this interesting line of study.}.  We study fluctuations of a static flow and obtain analytic expressions for the dispersion relations at third order in gradient expansion for shear and sound waves in a static reference frame, as well as the algebraic equations governing the dispersion relations for such modes in a boosted frame.

We applied the results of the general analysis in the study of the 
AdS$_4$/CFT$_{3}$ correspondence. We focused on the vector sector  
of the gravitational perturbations of the AdS$_4$ black-hole background written 
in terms of the RWZ variable.  
We solved the differential equation in the hydrodynamic limit using 
a double series expansion, in order to obtain the wavefunction 
in this regime. Additionally, the dispersion relations of the perturbations were
obtained up to fourth-order corrections in momentum. 
Next, we found the wavefunction for the RWZ variable of the scalar 
sector by using the SUSY relation, cf.~Eq.~\eqref{EqRWZScaToVec}.
In the sequence we mapped this solution into the 
corresponding EJL variable, which relates QN frequencies to the poles of 
retarded Green functions by imposing Dirichlet condition at the boundary, 
cf. Eq.~\eqref{EqEJLScaToSca}. The dispersion relations of this sector 
are obtained up to fourth-order in momentum, just like the vector sector. 
We also showed the agreement of the result obtained in the vector sector 
with results available in the literature. The same is true for the scalar sector:
the results obtained are in agreement with the known results up to third-order in 
momentum. However, so far there is no results reported in the literature 
considering the forth-order correction in momentum. Finally, we 
obtained the transport coefficients 
by comparing the dispersion relations for the 
gravitational perturbations of static black branes to the corresponding 
dispersion relations obtained from linear fluctuations of a 
relativistic conformal fluid. 
We realized that the transport coefficients arising from the 
third-order hydrodynamics do not depend on the temperature.

\section*{Acknowledgments}

L. A. H. M., S. M. D., and V. T. Z. thank financial support from 
Coordena\c{c}\~ao de Aperfei\c{c}oamento do Pessoal 
de N\'ivel Superior (CAPES, Brazil), Programa Nacional de P\'os-Doutorado and 
 Grant No.~88881.310352/2018-01.
A. S. M. thanks financial support from Funda\c{c}\~ao de Amparo
\`a Pesquisa do Estado da Bahia (FAPESB, Brazil).
V. T. Z. thanks financial support from Conselho 
Nacional de Desenvolvimento Cient\'\i fico
e Tecnol\'ogico (CNPq, Brazil), Grant No.~309609/2018-6.

\appendix

\section{Perturbative solution of the vector sector}
\label{Sec:AppendixA}
In this Appendix we are going to write details on the perturbative solution of the vector sector of 
the gravitational perturbation. In the forthcoming 
analysis we follow the ideas implemented in Refs.~\cite{Ge:2010yc,Brattan:2010pq}. 
By replacing Eq.~\eqref{EqPhiVec} in Eq.~\eqref{eqRWZ} 
we obtain the following differential equation
\noindent
\begin{equation}\label{EqF}
\begin{split}
&\left[-\frac{\mathfrak{q}^2}{f(u)}- \frac{\mathfrak{w}^2 \left(f'^2-9\right)}{9 f^2}
-\mathfrak{w} \left(\frac{i\, f''}{3 f}+\frac{2 i f'\,G'(u)}{3 f\,G(u)}\right)
+\frac{3 u}{f} +\frac{f'\, G'(u)}{f\,G(u)} +\frac{G''(u)}{G(u)}\right]F(u) \\
&+\left[\frac{2 G'(u)}{G(u)}+\left(1 -\frac{2\, i\, \mathfrak{w}}{3}\right)\frac{f'}{f}\right]F'(u)
+F''(u)=0.
\end{split}
\end{equation}
\noindent
In this equation we observe terms, in the coefficient of $F(u)$, which do not depend on the frequency 
and wavenumbers, so that we may separate these terms as an differential equation, 
\noindent
\begin{equation}\label{EqG}
\frac{3 u}{f}G(u) +\frac{f'}{f}G'(u)+G''(u)=0.
\end{equation}
\noindent
This is a smart way to split the problem in one piece depending on the holographic coordinate, $G(u)$,
and another depending on the coordinate, frequency and wavenumber, $F(u)$.
The solution to the differential Eq.~\eqref{EqG} used here is the simplest one, given 
by $G(u)=a_0\,u$, where $a_0$ is a constant.

On the other hand, to find the solutions for $F(u)$ we must replace Eq.~\eqref{EqExpansionF} into 
Eq.~\eqref{EqF} and rearrange the resulting differential 
equation to get differential equations for each one of the functions 
$F_1, F_2, \cdots$. For instance, the 
differential equation for $F_1$ is 
\noindent
\begin{equation}\label{eq:forFu}
\left(\frac{f'}{f}+\frac{2 G'(u)}{G(u)}\right)F_1'(u) 
+F_1''(u)=\left(\frac{i f''}{3 f}+\frac{2 i f'\, G'(u)}{3 f\,
G(u)}\right)F_0.
\end{equation}
\noindent
We do not write here the differential equations for the 
other functions: $F_2, F_3, F_4, F_5$, however, for completeness we list the 
solutions of such equations, for the functions $F_1, F_2,\cdots, F_5$, 
\noindent 
\begin{align*} 
F_1(u)=&\frac{i F_0}{18}\left(18+2\sqrt{3}\,\pi
-9\ln{3}\right)\\
&+iF_0\left(-\frac{1}{u}
-\frac{1}{\sqrt{3}}\arctan{\left[\frac{1+2\,u}{\sqrt{3}}\right]}
+\frac{1}{2}\ln\left[1+u+u^2\right]\right),\\ 
F_2(u)=&\frac{F_0}{3}\left(\frac{1}{u}-1\right),\\ 
F_3(u)=&\frac{F_0}{432}\bigg(72\, \text{Li}_2\left(
\frac{(-1)^{1/6} (1-u)}{\sqrt{3}}\right)
+72 \text{Li}_2\left(
\frac{(-1)^{5/6}
(u-1)}{\sqrt{3}}\right)\\
&-24\, i\, \sqrt{3}\, \text{Li}_2\left(e^{-\frac{2}{3} i \left(\pi 
-3 \arctan\left[\frac{2
u+1}{\sqrt{3}}\right]\right)}\right)
+72\, i\, \sqrt{3}\, \text{Li}_2\left(-e^{2 i \arctan\left[\frac{2 u+1}{\sqrt{3}}\right]}\right)\\
&-36 \ln^2\left(1-u^3\right)+72 (\ln{3}-3) \ln \left[u^2+u+1\right]\\&
+4 \sqrt{3} \pi  \bigg(\ln \left[\frac{-36 \left(2 \sqrt{3} u+\sqrt{3}+3
i\right)^2}{\left(\left(\sqrt{3}+i\right) u+\sqrt{3}-i\right)^4 \left(u^2+u+1\right)^3}\right]-4 \ln [1-u]\\
&-2 i \arctan\left[\frac{2u+1}{\sqrt{3}}\right]+3 \ln{3}\bigg)\\
&+36 \ln [1-u] \left(\ln \left[1-u^3\right]
+\ln \left[u^2+u+1\right]\right)+48 i \sqrt{3} \arctan\left[\frac{2 u+1}{\sqrt{3}}\right]^2\\
&+8 \sqrt{3} \bigg(6 \ln \left[\frac{1}{216} \left(3+i \sqrt{3}\right) (u-1) \left(2 \sqrt{3}
u+\sqrt{3}+3 i\right)^2\right]\\
&+5 i \pi +18\bigg) \arctan\left[\frac{2 u+1}{\sqrt{3}}\right]-6 i \sqrt{3} \pi  (\pi -8 i)-36 (\ln 3-6)
\ln 3\\
&+2 \left(-18 \psi ^{(1)}\left(\frac{1}{3}\right)+\left(12-\frac{4 i}{\sqrt{3}}\right) \pi ^2\right)\bigg),\\ 
\end{align*}
\begin{align*} 
F_4(u)=&\frac{iF_0}{18}\left(2\sqrt{3}\,\pi+3\ln{3}\right)
-\frac{iF_0}{54\,u}\bigg(9\ln{3}-2\sqrt{3}\,\pi -9(1-u)\ln{\left[1+u+u^2\right]}\\
&+6\sqrt{3}\,(1+3\,u)\arctan{\left[\frac{1+2\,u}{\sqrt{3}} \right]}\bigg),\\ 
F_5(u)=&\frac{F_0}{54} \left(-2 \sqrt{3}\, \pi-3\right) +\frac{F_0}{18}\left(\frac{1}{u}
+2 \sqrt{3}\, \arctan\left[\frac{2u+1}{\sqrt{3}}\right]\right),
\end{align*} 
\noindent
where $\text{Li}_2\left(\frac{(-1)^{5/6} (u-1)}{\sqrt{3}}\right)$ is the polylogarithm and 
$\psi ^{(1)}\left(\frac{1}{3}\right)$ is the polygamma function. It is worth mentioning that all
the integration constants were fixed by the requirement of regularity at the horizon (at $u=1$). 

In turn, the approximation of the wave function close to the 
boundary may be written as, c.f. Eq.~\eqref{EqVectorBound}, 
\noindent
\begin{equation}
\Phi_{\scriptscriptstyle{V}}(u)=
\widetilde{\Phi}_{\scriptscriptstyle{V}}
\left(1+\Pi_{\scriptscriptstyle{V}}\,u+\cdots\right),
\end{equation}
\noindent
where $\widetilde{\Phi}_{\scriptscriptstyle{V}}$ and 
$\Pi_{\scriptscriptstyle{V}}$ depend on the wavenumbers. As commented in Sec.~\ref{Sec:PerturbVectorSector}, the pole 
of $\Pi_{\scriptscriptstyle{V}}$ gives us the dispersion relation for the vector sector. The explicit form 
of $\Pi_{\scriptscriptstyle{V}}$ is
\noindent
\begin{equation}
\Pi_{\scriptscriptstyle{V}}(\mathfrak{w}, \mathfrak{q})=
\frac{d(\mathfrak{w}, \mathfrak{q})} {i\,\mathfrak{w}\left(
\mathfrak{q}^2\left(\sqrt{3}\,\pi-9\ln{3}\right)-54\right)
+3\,\mathfrak{q}^2\left(6+\mathfrak{q}^2\right)},
\end{equation}
\noindent
where $d(\mathfrak{w}, \mathfrak{q})$ is given by 
\noindent
\begin{equation}
\begin{split}
d(\mathfrak{w}, \mathfrak{q})=& 1 -\frac{\mathfrak{q}^2}{3} 
-\frac{1}{54} \left(3+\sqrt{3}\, \pi \right)\mathfrak{q}^4\\
&+\mathfrak{w} \bigg(\frac{i}{18}\left(18 +\sqrt{3}\, \pi -9 \ln{3}\right)
-\frac{i}{54}\mathfrak{q}^2 \left(-18+\sqrt{3}\, \pi -9 \ln{3}\right)\bigg)\\
&+\mathfrak{w}^2 \bigg(\frac{\psi ^{(1)}\left(\frac{1}{3}\right)}{12}
+\frac{1}{648} \bigg[36\, i\, \sqrt{3} \left(3\, \text{Li}_2\left(-\sqrt[3]{-1}\right)
-\text{Li}_2\left(\sqrt[3]{-1}\right)\right)\\
&+\pi  \left(-540+36 i \sqrt{3}+63\, \pi +7\,i\,\sqrt{3}\, \pi\right)-36 (\pi -18) \ln{2}\\
&-9 \ln{3} \left(-72+2\, \pi -2 \sqrt{3}\, \pi +18 \ln{2}+9 \ln{3}\right)\bigg]\bigg).
\end{split}
\end{equation}
\noindent

\section{Irreducible gradients not present in the gradient expansion}
\label{Sec:AppendixB}

Vector corrections to the energy-momentum tensor in the general case af non-conformal fluid can be ruled out by choosing Landau frame and, hence, they do not introduce new transport coefficients in addition to the set needed by scalar and tensor corrections. In the case of conformally symmetric fluids, the traceless nature of the energy-momentum tensor constrains all transport coefficients coming from conformal scalars to vanish. This is the reason why the list of irreducible conformal and non-conformal vectors and conformal scalars were not written in section \ref{Sec:ConformalFluids} and \ref{Sec:NonConformalFluids}. For completeness, these kinds of structures are listed below.

\subsection{Non-conformal vectors}
For ordinary non-conformal fluids we have at first order only one admissible transverse vector, namely, $\nabla_{_\perp}^\mu \ln s$.
To build second order vectors we can take gradients of a first order scalar, divergences of a first order tensor and also multiply two first order structures, obtaining the list
 \begin{multicols}{3}
\begin{enumerate}
 \item  $ \nabla_{\perp}^\mu \Theta$,
 \item $ \nabla_{\perp \nu}\sigma^{\mu\nu}- u^\mu\sigma^2 $,
 \item $\nabla_{\perp \nu}\Omega^{\mu\nu}-u^\mu\Omega^2  $,
 \item $    \Theta\nabla_{\perp }^\mu \ln s $,
 \item $  \sigma^{\mu\nu}\nabla_{\perp \nu} \ln s $,
 \item $  \Omega^{\mu\nu}\nabla_{\perp \nu} \ln s $,
  \item $ \Delta^{\mu\nu}u^\eta R_{\nu\eta}$.
\end{enumerate}
 \end{multicols}

The third order vectors are found by following the same strategy. We can take gradients of second order scalars, divergence of second order tensors or multiplying one second order with one third order structure. There is no special constraint in addition to those already presented above, and the complete list we find is
\begin{multicols}{3}
\begin{enumerate}
 \item  $\nabla_{_\perp}^\mu\Theta^2$,
 \item $\sigma^{\mu\nu}\nabla_{\perp \nu}\Theta$,
 \item $\Omega^{\mu\nu}\nabla_{\perp \nu}\Theta$,
  \item $\Theta (\nabla_{\perp \nu}\sigma^{\mu\nu}- u^\mu\sigma^2)$,
 \item $\Theta(\nabla_{\perp \nu}\Omega^{\mu\nu}-u^\mu\Omega^2)$ ,
 \item $ \Theta^2\nabla_{_\perp}^\mu \ln s$,
 \item $  \Theta\sigma^{\mu\nu}\nabla_{_\perp \nu} \ln s$,
 \item $   \Theta\Omega^{\mu\nu}\nabla_{_\perp \nu}\ln s$,
 \item $ D\Theta \nabla_{\perp}^\mu\ln s $,
 \item $\nabla_{\perp}^\mu D\Theta $,
 \item $ \nabla_{_\perp}^\mu \Omega^2$ ,
 \item $\Delta^{\mu \eta}\sigma^{\alpha\beta}\nabla_{_\perp \beta}\sigma_{\alpha\eta}$,
  \item $\Delta^{\mu \eta}\Omega^{\alpha\beta}\nabla_{_\perp \beta}\Omega_{\alpha\eta}$,
 \item $\Omega^2 \nabla_{_\perp}^\mu\ln s $,
 \item $\sigma^2 \nabla_{_\perp}^\mu\ln s $,
 \item $ \sigma^{\mu\nu}\sigma_{\nu\eta}\nabla_{_\perp}^\eta\ln s$,
 \item $ \Omega^{\mu\nu}\Omega_{\nu\eta}\nabla_{_\perp}^\eta\ln s$,
  \item $ \sigma^{\mu\nu}\Omega_{\nu\eta}\nabla_{_\perp}^\eta\ln s$,
   \item $ \Omega^{\mu\nu}\sigma_{\nu\eta}\nabla_{_\perp}^\eta\ln s$,
 \item $  (\nabla_{_\perp}\ln s)^2 \nabla_{_\perp}^\mu\ln s $,
  \item $  \nabla_{_\perp}^\mu (\nabla_{_\perp}\ln s)^2$,
  \item $  \nabla_{_\perp}^\mu (u^\alpha u^\beta R_{\alpha\beta})$,
    \item $  u^\alpha u^\beta R_{\alpha\beta}\nabla_{_\perp}^\mu\ln s$,
  \item $\Delta^{\mu\eta}u^\alpha u^\beta \nabla^\gamma R_{\eta\alpha\beta\gamma}$,
  \item $  \Delta^{\mu\nu} R_{\nu\beta}\nabla_{_\perp}^\beta \ln s$,
  \item $  \nabla_{_\perp}^\mu R$,
  \item $  R \nabla_{_\perp}^\mu \ln s$,
  \item $\Delta^{\mu\eta}u^\alpha u^\beta   R_{\eta\alpha\beta\gamma}\nabla_{_\perp}^\gamma\ln s$,
  \item $\Delta^{\mu\eta}\sigma^{\alpha \beta}u^\gamma  R_{\eta\alpha\beta\gamma}$,
  \item $\Delta^{\mu\eta}\Omega^{\alpha \beta}u^\gamma  R_{\eta\alpha\beta\gamma}$,
  \item $\sigma^{\mu\nu}u^\eta R_{\nu\eta}$,
    \item $\Omega^{\mu\nu}u^\eta R_{\nu\eta}$,
      \item $\Delta^{\mu\nu}u^\eta R_{\nu\eta}$.
      \item  $\nabla_{\perp \mu}\sigma^2$.
    
  \end{enumerate}
\end{multicols}
 This list presents 34 independent vectors and differs from the one presented in \cite{Grozdanov:2015kqa}, where  there are only 28 vectors and we can note, for example, that $\nabla_\perp^\mu R$ is missing there. Differently from the scalars and tensors, we found a larger list of independent vectors at third order in gradient expansion.

\subsection{Conformal vectors}  

The conformal symmetry reduces the number of possible structures in comparison to the non-conformal fluid. At first order there is no conformal transverse vectors, and the only first order vectorial structure is in fact the connection defined in eq.(\ref{connection}). 

To get a second order conformal vectors we can take divergence of a first order tensor (presented in section \ref{sec:order1and2}) and also project the second order tensor  along the flow. At second order we have four  conformal vectors, namely,
 \begin{multicols}{2}
\begin{enumerate}
 \item  $u_\sigma\mathcal{F}^{\mu\sigma}$,
 \item $\Delta^\mu_{~\nu}u_\eta\mathcal{R}^{\nu\eta} $,
 \item $\Delta^\mu_{~\nu} \mathcal{D}_\eta \sigma^{\nu\eta} $,
 \item $  \Delta^\mu_{~\nu} \mathcal{D}_\eta \Omega^{\nu\eta} $.
\end{enumerate}
 \end{multicols}
  It is remarkable that the first vector in the list has a nice physical meaning. It is
  the accelerating force on a test particle due to the presence of the tensor field of the Weyl connection, being  analogous to the Lorentz force due to the Faraday-Maxwell electromagnetic tensor field. The vectors $u_\sigma\mathcal{F}^{\mu\sigma},~\Delta^\mu_{~\nu}u_\eta\mathcal{R}^{\nu\eta}$ we present here do not appear in previous works.  In particular, it is staded in Ref.~ \cite{Bhattacharyya:2008jc} that there are only two second order conformal vectors. This increment of two vectors follows from implementing Weyl covariance  through the Weyl connection, generating the conformal covariant tensor $\mathcal{F}^{\mu\nu}$ and allowing for a second order Weyl covariant vector associated to the Ricci tensor. 
  
  Third order conformal vectors are obtained by the same strategy resulting in the following list
\begin{multicols}{3}
\begin{enumerate}
 \item $\Delta^\mu_{~\nu}\mathcal{D}^\nu\sigma^2$,
 
  \item $\Delta^\mu_{~\nu}\mathcal{D}^\nu\Omega^2$,
  
  \item $\Delta^\mu_{~\nu}\mathcal{D}^\beta(\sigma_\eta^{~\nu}\sigma_\beta^{~\eta})$,
  
  \item $\Delta^\mu_{~\nu}\mathcal{D}^\beta(\sigma_\eta^{~\nu}\Omega_\beta^{~\eta})$,
  
    \item $\Delta^\mu_{~\nu}\mathcal{D}^\beta(\Omega_\eta^{~\nu}\Omega_\beta^{~\eta})$,
    
      \item $\Delta^\mu_{~\nu}\mathcal{D}_\beta\mathcal{R}^{\nu\beta}$,
      
      \item  $\Delta^\mu_{~\nu}\sigma^{\nu\beta}\mathcal{R}^{\beta\eta}u_\eta$,
      
      \item  $\Delta^\mu_{~\nu}\Omega^{\nu\beta}\mathcal{R}^{\beta\eta}u_\eta$,
      
      \item  $\Delta^{\mu\eta}\Omega^{\nu\beta}u^\delta\mathcal{R}_{\eta\nu\delta\beta}$,
      
        \item  $\Delta^{\mu\eta}\sigma^{\nu\beta}u^\delta\mathcal{R}_{\eta\nu\delta\beta}$,
      
        \item  $\Delta^{\mu\nu}u^\beta u^\delta\mathcal{D}_\eta  \mathcal{R}^\eta_{~\nu\delta\beta}$.
\end{enumerate}
\end{multicols}

\subsection{Conformal scalars}

 There are no first order conformal scalars.
 
 Considering second order conformal scalars we have the scalar product of first order structures and the scalar projections of the conformal Riemann tensor, namely, $\sigma^2$, $ \Omega^2,~\mathcal{R}$, and $u^\mu u^\nu\mathcal{R}_{\mu\nu}$, respectively. Notice that the field tensor $\mathcal{F}^{\mu\nu}$ is of second order but, due to its anti-symmetric character, it is traceless and  we have the algebraic relation $u^\mu u^\nu \mathcal{R}_{\mu\nu} = \sigma^2+\Omega^2$, which allows us to remove also the structure $u^\mu u^\nu \mathcal{R}_{\mu\nu}$ from the list. So, there are only three independent second order conformal scalars,
\begin{multicols}{3}
 \begin{enumerate}
\item $ \mathfrak{P}^{^{2nd}}_1=\Omega^2$,\item $ \mathfrak{P}^{^{2nd}}_2=\sigma^2$, \item $\mathfrak{P}^{^{2nd}}_3=\mathcal{R}$.
 \end{enumerate}
\end{multicols}

Third order conformal scalars can be made by products (contractions) of first-order and second-order structures, and also by taking total derivatives of second order scalars.  We have schematically  
$\sigma\Omega^2 \sim   u^\mu \mathcal{D}_\mu\Omega^2+\mathcal{D}^2\Omega,$ $u_\mu\mathcal{D}_\nu \mathcal{R}^{\mu \nu} \sim u^\mu\mathcal{D}_\mu \mathcal{R}$, and
$\Omega_{\mu\nu}\mathcal{R}^{\mu\nu}= \Omega_{\mu\nu}\mathcal{F}^{\mu\nu} \sim \mathcal{D}_\mu\mathcal{D}_\nu\Omega^{\mu\nu}$, which eliminates four scalars. We also have 
$ \Omega\mathcal{F} \sim \sigma\Omega^2+ u^\mu\mathcal{D}_\mu \Omega^2$, and
$\sigma\Omega^2 \sim u^\mu\mathcal{D}_\mu \Omega^2 + \mathcal{D}_a\mathcal{D}_b\Omega^{ab}$, which eliminates two more scalars, and the anty-symmetry of $\Omega$ gives that $\Omega^3$ vanishes identically. Equation (A.1) in \cite{Loganayagam:2008is} allows to express $\Omega^{\mu\nu}u^\alpha u^\beta\mathcal{R}_{\mu\alpha\nu\beta}$ and $\sigma^{\mu\nu}u^\alpha u^\beta\mathcal{R}_{\mu\alpha\nu\beta}$ as a combination of the other third order scalars, 
eliminating the two scalars from the list of third-order independent scalar structure. Finally we end up with a list of six conformal third order scalars, namely, 
\begin{multicols}{3}
\begin{enumerate}
 \item  $\mathfrak{P}^{^{3rd}}_1= \Omega_{\mu\nu}\mathcal{F}^{\mu\nu}$, \item $\mathfrak{P}^{^{3rd}}_2=u^\mu\mathcal{D}_\mu\sigma^2 $,
\item $ \mathfrak{P}^{^{3rd}}_3=\sigma^{\mu\nu}\sigma_{\nu\eta}\sigma^\eta_{~\mu}$, \item $ \mathfrak{P}^{^{3rd}}_4= \mathcal{D}_\mu\mathcal{D}_\nu\sigma^{\mu\nu}$, \item $\mathfrak{P}^{^{3rd}}_5= u^\mu\mathcal{D}_\mu\mathcal{R}$, \item $\mathfrak{P}^{^{3rd}}_6= \sigma_{\mu\nu}\mathcal{R}^{\mu\nu}$.
\end{enumerate}
\end{multicols}

\section{The ordering  of transverse derivatives doesn't matter }
\label{Sec:AppendixC}
In the gradient expansion of an uncharged fluid the ordering of transverse derivatives is irrelevant, a change in the order of two successive transverse derivatives does not give rise to new transport coefficients. Here we show how this mechanism works for third order gradients. For the case of scalars we display the complete set of equivalences while for the tensors we display one case that illustrates how the equivalence works. The remaining redundancies of third order tensors are obtained in a straightforward manner.

The third order scalars in the sets $\{ \nabla_{\perp \mu}\nabla_{\perp}^\mu\nabla_{\perp \nu} u^\nu, 
\nabla_{\perp \mu}\nabla_{\perp \nu}\nabla_{\perp}^\mu u^\nu, \nabla_{\perp \mu}
\nabla_{\perp \nu}\nabla_{\perp}^\mu u^\nu\}$,  $\{\nabla_{\perp}^\mu\ln s \nabla_{\perp \mu}\nabla_{\perp \nu} u^\nu, \nabla_{\perp}^\mu\ln s \nabla_{\perp \nu}\nabla_{\perp \mu} u^\nu\}$, and $\{u^\mu\nabla_\nu R_\mu^{~\nu},~ u^\mu \nabla_\mu R\}$
 are presented in Appendix D of Ref.~\cite{Grozdanov:2015kqa}  as independent gradients. However, they are not independent, each one of these sets contains only one independent gradient. In order to reveal these redundancies we compute the differences between each pair of scalars obtaining a linear combination of the remaining  scalars  of the list \cite{Grozdanov:2015kqa}. Let us consider three cases explicitly here.
 \begin{align} 
   &\left(\nabla_{\perp}^\mu\ln s\right)\Big( \nabla_{\perp \mu}\nabla_{\perp \nu} u^\nu - \nabla_{\perp \nu}\nabla_{\perp \mu} u^\nu \Big) \nonumber \\
  &= \left(\nabla_{\perp}^\mu\ln s\right)\Big(\nabla_{\perp \mu} u_\nu D u^\nu - \nabla_{\perp \nu} u_\mu D u^\nu +\Delta_\mu^\sigma\Delta_\nu^\gamma\left[ \nabla_\sigma, \nabla_\gamma\right] u^\nu \Big) \nonumber \\
&= - \left(\nabla_\perp^\mu \ln s\right) R_{\mu\nu}u^\nu,   \\
&\nonumber \\
&\nabla_{\perp \mu}\Big(\nabla_{\perp}^\mu\nabla_{\perp \nu} u^\nu - \nabla_{\perp \nu}\nabla_{\perp}^\mu u^\nu\Big) \nonumber \\   
&=   \nabla_{\perp \mu}\Big(\nabla_{\perp}^\mu u_\nu D u^\nu - u^\mu \nabla_{\perp \nu} u^\gamma \nabla_{\perp \gamma} u^\nu   - \nabla_{\perp \nu } u^\mu D u^\nu  +\Delta_\mu^\sigma\Delta_\nu^\gamma[ \nabla_\sigma, \nabla_\gamma] u^\nu  \Big) \nonumber \\ 
&=  \nabla_{\perp \mu}\Big(c_s^2\left[\nabla_{\perp}^\nu u^\mu-\nabla_{\perp}^\mu u^\nu\right]\nabla_{\perp \nu }\ln s - u^\mu \nabla_{\perp \nu} u^\gamma \nabla_{\perp \gamma} u^\nu   + R^\mu_\nu u^\nu + u^\mu u^\nu u^\alpha R_{\nu \alpha}\Big) \nonumber \\
&= c_s^2 \nabla_{\perp}^\mu u^\nu \Big( \nabla_{\perp \nu } \nabla_{\perp \mu }\ln s - \nabla_{\perp \mu } \nabla_{\perp \nu }\ln s\Big) +c_s^2 \left(\nabla_{\perp \nu }\ln s\right)\nabla_{\perp \mu }\Big(\nabla_{\perp}^\nu u^\mu - \nabla_{\perp}^\mu u^\nu\Big)  \nonumber \\ 
~&-  \nabla_{\perp \mu }u^\mu \nabla_{\perp \nu} u^\gamma \nabla_{\perp \gamma} u^\nu + R_{\mu\nu}\nabla_{\perp}^\mu u^\nu + u^\mu \nabla_{\perp}^\nu R_{\mu\nu}  + u^\alpha u^\beta R_{\alpha \beta} \nabla_{\perp \mu} u^\mu \nonumber \\ 
&= c_s^2 \nabla_{\perp \mu }u^\mu\Big( \nabla_{\perp \nu} u_\gamma \nabla_{\perp}^\nu u^\gamma  -  \nabla_{\perp \nu} u_\gamma \nabla_{\perp}^\gamma u^\nu\Big) + c_s^2\nabla_{\perp \nu }\ln s  \Big(\nabla_{\perp}^\nu u^\mu  -\nabla_{\perp \mu }\nabla_{\perp}^\mu u^\nu \Big) \nonumber \\ ~&-\nabla_{\perp \mu }u^\mu \nabla_{\perp \nu} u_\gamma \nabla_{\perp}^\gamma u^\nu +  R_{\mu\nu}\nabla_{\perp}^\mu u^\nu + u^\mu \nabla_\mu R + u^\mu u^\nu u ^\alpha \nabla_\alpha R_{\mu\nu} + u^\alpha u^\beta R_{\alpha \beta} \nabla_{\perp \mu} u^\mu, \\
 & \nabla_{\perp \mu}\nabla_{\perp \nu}\nabla_{\perp}^\mu u^\nu - \nabla_{\perp \nu} \nabla_{\perp \mu}\nabla_{\perp}^\mu u^\nu  \nonumber \\
 & = \nabla_{\perp \mu }u_\nu D \nabla_{\perp }^\mu u^\nu +\nabla_{\perp \mu }u^\gamma u_\nu \nabla_{\perp \gamma } \nabla_{\perp}^\mu u^\nu -  \nabla_{\perp \nu }u_\mu D \nabla_{\perp }^\mu u^\nu - \nabla_{\perp \nu }u^\gamma u_\mu \nabla_{\perp \gamma } \nabla_{\perp}^\mu u^\nu \nonumber \\
 &+ \Delta_\mu^{~\sigma}\Delta_\nu^{~\gamma}[\nabla_\sigma, \nabla_\gamma] \nabla_{\perp}^\mu u^\nu \nonumber \\
 &= - \nabla_{\perp \mu} u_\nu \nabla_{\perp}^\mu u^\gamma \nabla_{\perp \gamma} u^\nu 
 - c_s^2\nabla_{\perp \mu} u_\nu \nabla_{\perp}^\mu  \nabla_{\perp}^\nu \ln s + 
 u^\gamma u^\sigma R^{\nu ~\mu}_{~\gamma ~\sigma} \nabla_{\perp \mu} u_\nu  + \nabla_{\perp \mu} u_\nu Du^\mu Du^\nu \nonumber \\
& - \nabla_{\perp \mu} u^\gamma \nabla_{\perp \gamma} u_\nu\nabla_{\perp}^\mu u^\nu
 + \nabla_{\perp \nu} u^\gamma \nabla_{\perp \gamma} u_\mu \nabla_{\perp}^\mu u^\nu + c_s^2c_s^2\nabla_{\perp \nu} u_\mu \nabla_{\perp}^\mu  \nabla_{\perp}^\nu \ln s  -  u^\gamma u^\sigma R^{\mu ~\nu}_{~\gamma ~\sigma} \nabla_{\perp \mu} u_\nu \nonumber \\
 &-\nabla_{\perp \nu} u_\mu Du^\mu Du^\nu +  \nabla_{\perp \mu} u^\gamma \nabla_{\perp \gamma} u_\nu\nabla_{\perp}^\mu u^\nu
 + \Delta_\mu^{~\sigma}\Delta_\nu^{~\gamma}( R^\mu_{~\sigma\gamma\alpha} \nabla_{\perp}^\alpha u^\nu + 
 R^\nu_{~\sigma\gamma\alpha} \nabla_{\perp}^\mu u^\alpha)  \nonumber \\
 &= - \nabla_{\perp \mu} u_\nu \nabla_{\perp}^\mu u^\gamma \nabla_{\perp \gamma} u^\nu   
 + c_s^2( \nabla_{\perp \alpha} u^\alpha \nabla_{\perp}^\mu u^\nu \nabla_{\perp \mu} u_\nu  -  \nabla_{\perp \alpha} u^\alpha \nabla_{\perp}^\mu u^\nu \nabla_{\perp \nu} u_\mu) \nonumber \\ 
 &+ R_{\mu \nu}\nabla_{\perp}^\mu u^\nu   + u^\gamma u^\nu \nabla_{\perp}^\mu u^\alpha  R_{\nu \mu\gamma\alpha}.
 \end{align}  
Within the list of third order scalars presented in \cite{Grozdanov:2015kqa}, the above equations translates in
\begin{align} 
 S_7 &= S_8 - S_{19}, \nonumber \\
S_1 &= S_2 + c_s^2(S_{12} + S_8 - S_6) - (1+c_s^2)S_{13} + S_{17} + S_{18} + S_{21} + S_{22}, \nonumber \\
  S_2 &= S_3 +c_s^2(S_{12}-S_{13}) + S_{21} + S_{23}. \label{eq:equivs}
\end{align} 
Moreover, the Bianchi identity (i.e., the fact that the Einstein tensor is divergence free) provides $S_{16} = S_{17}$. From this equality and from the three relations \eqref{eq:equivs}, we then conclude that four terms from the list of 23 scalars presented in \cite{Grozdanov:2015kqa} can be eliminated. For instance, $S_1,\, S_2,\, S_7,$, and $ S_{16}$ can be expressed as linear combinations of $S_3$, $S_6$,  $S_8$, $S_{12}$, $S_{13}$, $S_{17}$, $S_{18}$, $S_{19}$, $S_{21}$, $S_{22},$ and $S_{23}$. For this reason there are only 19 independent third order scalars in the gradient expansion of a non-conformal fluid.   
These additional equivalences appear as a direct consequence of the facts that it is possible to replace the covariant derivative by its transverse version when acting on $u^\mu$ or $\ln s$, and that the commutation of two covariant derivatives gives rise to Riemann tensors only. At the end of the day,  we have that in the gradient expansion the equivalence $\nabla_{\perp \mu} \nabla_{\perp \nu} \sim \nabla_{\perp \nu} \nabla_{\perp \mu}$ holds. 

For the transverse and trace-less tensors one faces the same situation: the change in the order of transverse derivatives does not give raise to new independent tensors in the gradient expansion. 
The explicit equivalences are obtained in a similar way as done above for the third order scalars. To  illustrate the procedure we show here explicitly one case: the equivalence between the tensors $\mathcal{T}_1$ and $\mathcal{T}_2$ presented in Appendix C of Ref. \cite{Grozdanov:2015kqa}.   
In order to do that we compute the difference between  
$\nabla_{\perp \alpha}\nabla_{\perp}^\alpha\nabla_{\perp}^{ \mu} u^{\nu}$ and 
$\nabla_{\perp \alpha}\nabla_{\perp}^{\mu} \nabla_{\perp}^\alpha u^{\nu}$
by taking the TST part and expressing the result as a linear combination of the remaining third order TST tensors. Notice that we omit the therms whose TST parts vanish.
\begin{align}
&\nabla_{\perp \alpha}\nabla_{\perp}^\alpha\nabla_{\perp}^{ \mu} u^{\nu} - \nabla_{\perp \alpha}\nabla_{\perp}^{\mu} \nabla_{\perp}^\alpha u^{\nu} \nonumber \\ 
&= \nabla_{\perp \alpha}\Big( u^\sigma \nabla_{\perp}^\alpha u^\mu \nabla_{\sigma}u^\nu + u^\mu \nabla_{\perp}^\alpha u^\sigma \nabla_{\sigma}u^\nu 
-u^\alpha \nabla_{\perp}^\mu u^\sigma \nabla_{ \sigma}u^\nu -  u^\sigma \nabla_{\perp}^\mu u^\alpha\nabla_{ \sigma}u^\nu  \nonumber \\
& + \Delta^{\alpha\beta}\Delta^{\mu\sigma}[\nabla_\beta, \nabla_\sigma]u^\nu\Big) \nonumber \\ 
&= \nabla_{\perp \alpha}\nabla_{\perp}^\alpha u^\mu Du^\nu + \nabla_{\perp}^\alpha u^\mu \nabla_{\perp \alpha} Du^\nu +  \nabla_{\perp \alpha} u^\mu \nabla_{\perp}^\alpha u^\sigma  \nabla_{\perp \sigma} u^\nu
-  \nabla_{\perp\alpha} u^\alpha  \nabla_{\perp}^\mu u^\sigma  \nabla_{\perp \sigma} u^\nu \nonumber \\
& -  \nabla_{\perp \alpha}  \nabla_{\perp}^\mu u^\alpha Du^\nu 
+ u^\beta u^\sigma R^{\nu ~\mu}_{~\beta ~\sigma}\nabla_{\perp \alpha} u^\alpha +
u^\sigma \nabla_{\perp}^\beta R^{\nu ~\mu}_{~\beta ~\sigma} +R^{\nu ~\mu}_{~\beta ~\sigma}\nabla_{\perp}^\beta u^\sigma.
\end{align}
Therefore, the TST part gives
\begin{align}
 &\nabla_{\perp \alpha}\nabla_{\perp}^\alpha\nabla_{\perp}^{\langle \mu} u^{\nu\rangle} - \nabla_{\perp \alpha}\nabla_{\perp}^{\langle \mu} \nabla_{\perp}^\alpha u^{\nu\rangle} \nonumber \\ 
 &= c_s^2(  \nabla_{\perp \alpha} \nabla_{\perp}^{\langle \mu} u^\alpha \nabla_{\perp}^{\nu\rangle}\ln s -  \nabla_{\perp \alpha}\nabla_{\perp}^\alpha u^{\langle \mu} \nabla_{\perp}^{\nu\rangle}\ln s - \nabla_{\perp}^\alpha u^{\langle \mu}  \nabla_{\perp \alpha}\nabla_{\perp}^{\nu\rangle}\ln s) \nonumber \\
 &+  \nabla_{\perp}^\alpha u^\sigma  \nabla_{\perp \alpha} u^{\langle \mu}\nabla_{\perp \sigma} u^{\nu\rangle}
 - \nabla_{\perp \alpha} u^\alpha  \nabla_{\perp}^\sigma u^{\langle \mu}\nabla_{\perp \sigma} u^{\nu\rangle} -\nabla_{\perp \alpha} u^\alpha u^\beta u^\sigma R^{~\langle \mu\nu\rangle}_{\beta~~~\sigma} \nonumber \\ 
 &- u^\sigma \nabla^\beta R^{~\langle \mu\nu\rangle}_{\beta~~~\sigma} - u^\sigma u^\beta u^\gamma \nabla_\gamma R^{~\langle \mu\nu\rangle}_{\beta~~~\sigma} - \nabla_\perp^\beta u^\sigma R^{~\langle \mu\nu\rangle}_{\beta~~~\sigma}.
\end{align} 
According to the notation used in Appendix C of Ref.  \cite{Grozdanov:2015kqa}, we have just shown that 
\begin{equation}
 \mathcal{T}_1 = \mathcal{T}_2 + c_s^2(\mathcal{T}_{14}-\mathcal{T}_8-\mathcal{T}_{16}) - \mathcal{T}_{22} + \mathcal{T}_{26} - \mathcal{T}_{31}-\mathcal{T}_{34} - \mathcal{T}_{41} - \mathcal{T}_{45}.
\end{equation}

Other equivalences may be verified in a similar way as in the cases we worked out in this appendix.

\end{document}